\documentclass[letterpaper,twocolumn,10pt, screen]{article}
\usepackage{usenix-2020-09}

\usepackage{tikz}
\usepackage{amsmath}
\usepackage{xspace}
\usepackage{enumitem}

\usepackage{filecontents}

\usepackage{pifont}
\usepackage{txfonts}
\usepackage{array, boldline, makecell}
\usepackage{circledsteps}
\usepackage{color}

\usepackage{booktabs}
\usepackage{caption}
\usepackage{multirow}
\usepackage{subfig}
\usepackage{url}
\usepackage{textcomp}

\usepackage{blindtext}
\usepackage{authblk}
\newcommand{\sys}{PolarStore\xspace}

\newcommand{\dev}{PolarCSD}

\newcommand{\mycaption}[3]{{\caption{\label{#1}{\textbf{#2.} }{ #3}}}}

\newcommand{\cons}[1]{\textcolor{purple}{#1}}

\newcommand{\dbname}{PolarDB\xspace}
\usepackage{tikz,lipsum}
\usepackage[most]{tcolorbox}
\newtcolorbox{myframe}[2][]{%
	colback=gray!10,
	enhanced,colframe=black,coltitle=black,
	sharp corners,boxrule=0.4pt,
	fonttitle=\bfseries,
	attach boxed title to top left={yshift=-0.3\baselineskip-0.4pt,xshift=2mm},
	boxed title style={tile,size=tight,width=fit to content,left=0.5mm,right=0.5mm,
		colback=white,before upper=\strut},
	title={\parbox{0.95\textwidth}{#2}},#1
    before upper={\setlength{\parindent}{1em}}  
}
\tcbuselibrary{listings,breakable}

\newcommand{\myCircled}[2][]{\Circled[fill color=black, inner color=white, #1]{\sffamily {\scriptsize #2}}}

\usepackage[linesnumbered,ruled,vlined]{algorithm2e}

\usepackage{titlesec}
\titlespacing*{\section}{0pt}{2.5ex}{0.8ex}
\titlespacing*{\subsection}{0pt}{0.6ex}{0.2ex}
\titlespacing*{\subsubsection}{0pt}{0.2ex}{0ex}

\begin{document}


\date{}

\title{\sys: High-Performance Data Compression \\for Large-Scale Cloud-Native Databases}



\renewcommand*{\Authsep}{\quad}
\renewcommand*{\Authands}{\quad}
\author[]{\rm Qingda Hu}
\author[]{\rm Xinjun Yang}
\author[]{\rm Feifei Li}
\author[]{\rm Junru Li\thanks{Junru Li is the corresponding
author (rusuo.ljr@alibaba-inc.com).}}
\author[]{\rm Ya Lin}
\author[]{\rm Yuqi Zhou}
\author[]{\rm Yicong Zhu}
\author[]{\rm Junwei Zhang}
\author[]{\rm Rongbiao Xie}
\author[]{\rm Ling Zhou}
\author[]{\rm Bin Wu}
\author[]{\rm Wenchao Zhou}
\affil[]{\textit{Alibaba Cloud Computing}}

\maketitle
\begin{abstract}

\nocite{polarstore}
In recent years, resource elasticity and cost optimization have become essential for RDBMSs. While cloud-native RDBMSs provide elastic computing resources via disaggregated computing and storage, storage costs remain a critical user concern. Consequently, data compression emerges as an effective strategy to reduce storage costs. However, existing compression approaches in RDBMSs present a stark trade-off: software-based approaches incur significant performance overheads, while hardware-based alternatives lack the flexibility required for diverse database workloads.
In this paper, we present PolarStore, a compressed shared storage system for cloud-native RDBMSs. PolarStore employs a dual-layer compression mechanism that combines in-storage compression in PolarCSD hardware with lightweight compression in software. This design leverages the strengths of both approaches. PolarStore also incorporates database-oriented optimizations to maintain high performance on critical I/O paths. Drawing from large-scale deployment experiences, we also introduce hardware improvements for PolarCSD to ensure host-level stability and propose a compression-aware scheduling scheme to improve cluster-level space efficiency. PolarStore is currently deployed on thousands of storage servers within PolarDB, managing over 100\,PB of data. It achieves a compression ratio of 3.55 and reduces storage costs by approximately 60\%. Remarkably, these savings are achieved while maintaining performance comparable to uncompressed clusters.

\end{abstract}
\pagenumbering{gobble}

\section{Introduction}
Relational Database Management Systems (RDBMSs) are fundamental components of modern information technology infrastructure. In recent years, a growing number of applications have adopted cloud computing to address demands of resource elasticity and on-demand usage. Consequently, cloud service providers have developed comprehensive RDBMS solutions, such as AWS Aurora~\cite{DBLP:conf/sigmod/VerbitskiGSBGMK17}, Azure Hyperscale~\cite{DBLP:conf/sigmod/AntonopoulosBDS19}, and our production system, Alibaba PolarDB~\cite{feifeili, cloudjump, DBLP:journals/pvldb/CaoLWCZZWM18, cloudjump2}.
With the rapid expansion of data stored in the cloud, storage costs have become a significant concern for users. 	

To address storage costs, data compression techniques have emerged as an intuitive solution~\cite{TerseCades,QATFS,compressimage,mao2017elastic,cockshott1998data,iyer1994data,DBLP:conf/vldb/PossP03,gao2024revisiting,harnik2013zip, harnik2014fast}. Data compression approaches can be categorized into two main categories: software-based and hardware-based compression. Software-based compression utilizes CPU resources to execute compression algorithms and provide complex space management, while hardware-based compression offloads the compression tasks to specialized hardware.
However, implementing data compression to achieve high space utilization in large-scale RDBMSs presents several challenges. 

\noindent
\textbf{Challenge\#1: performance overhead of software-based compression.} 
Modern RDBMS business scenarios, such as online e-commerce and real-time financial transactions, demand low I/O latency. 
However, software-based compression faces not only computational overhead from compression operations but also a fundamental challenge in managing the mapping between original and compressed data. Since the compressed data size varies with content, systems must maintain an index to locate compressed data and handle size changes during updates. This indexing mechanism presents a critical trade-off: while fine-grained indexing could significantly improve space utilization, it introduces complex management overhead that impacts performance. This trade-off manifests differently across various RDBMS architectures: B\texttt{+}-Tree-based systems suffer from inherent space fragmentation due to 4KB block alignment~\cite{tablecompr,pagecompr}, while systems based on Log-Structured Merge-Trees (LSM-Tree) achieve a more compact data layout but incur substantial overhead from garbage collection~\cite{matsunobu2020myrocks,huang2020tidb,rocksDBcompr,levelcompr, yang2022oceanbase}. 
Therefore, achieving both high space efficiency and low I/O latency remains a significant challenge.


\noindent
\textbf{Challenge\#2: limited flexibility of hardware-based compression.} 
Given the performance overhead of software-based compression, researchers and industry practitioners have turned to hardware-based solutions~\cite{DBLP:conf/hotstorage/ZhengCLWLPSZ020, BreathingNewLife, chiosa2022hardware, DBLP:conf/fast/QiaoCZLL022, mircosoft, qat}.
However, these approaches introduce new challenges in terms of flexibility.
In-storage compression uses computational storage drives (CSDs)~\cite{DBLP:conf/hotstorage/ZhengCLWLPSZ020, BreathingNewLife}, which integrate computational capabilities to offload compression tasks to storage devices.
However, CSDs are constrained by fixed 4KB input sizes due to NVMe compatibility requirements and compression algorithms that cannot be modified after production. Similarly, PCIe-attached FPGA~\cite{mircosoft} or CPU-based accelerators~\cite{qat} are also limited to fixed algorithms. 
These limitations restrict the ability of RDBMSs to adapt key compression parameters (i.e., compression algorithms and input sizes) for diverse workload patterns. 
While some data requires real-time processing with minimal latency, other infrequently accessed data could benefit from more complex algorithms with larger input sizes to achieve higher compression ratios.
Therefore, it is beneficial to provide flexible compression solutions that can optimize storage efficiency for cold data while meeting the low-latency requirements of latency-sensitive workloads.

To address these challenges, we propose \sys, a storage system for RDBMSs that
co-designs hardware and software to achieve both high space utilization and low I/O latency. First, \sys implements a \textit{dual-layer compression mechanism} that processes data in two stages: 
the software layer compresses data into 4\,KB-aligned blocks, maintaining simple index management at the software level and providing flexible compression parameters (i.e., input sizes and algorithms), 
while the hardware layer, \dev, further compresses these blocks, leveraging the existing flash translation layer (FTL) to achieve byte-granularity indexing without additional software overhead. 
Second, we introduce \textit{several DB-oriented optimizations} to overcome the compression overhead. These optimizations target two critical I/O paths that directly impact user-perceived latency: redo log writes during transaction commits and page reads upon in-memory buffer pool misses.

\noindent
\textbf{Challenge\#3: stability and scalability of compression in large-scale deployment.} Deploying data compression at scale in RDBMSs presents new operational challenges.
At the host level, each server is equipped with 10\textasciitilde12 \dev~devices, and resource contention (e.g., CPU and memory) or faults from software drivers can lead to host-level failures and performance fluctuations. 
At the cluster level, varying compression ratios across different users' data make it challenging to balance logical and physical space among storage nodes. To address these deployment challenges, we redesign the hardware based on our deployment experiences and implement a compression-aware scheduling mechanism to balance compression ratios across storage nodes, thereby ensuring both host-level stability and cluster-level space efficiency.

\sys is deployed across thousands of storage servers and numerous clusters within \dbname~\footnote{See official document~\cite{polarstore} for details on enabling compression feature.}.
The total storage capacity has surpassed 100\,PB, achieving approximately 60\% storage cost reduction with a compression ratio of 3.55.
Comparative experiments with the clusters without data compression demonstrate that \sys maintains high performance with negligible degradation.

In summary, this paper makes the following contributions:
\begin{itemize}
    \item \sys, a shared storage system that leverages a hardware-software co-design for efficient compression. 
    \item A set of DB-oriented techniques that achieve high space efficiency without sacrificing the low I/O latency of critical operations.
    \item Valuable insights gained from the large-scale deployment in both CSD hardware design and cluster management, along with some practical solutions to address the encountered challenges.
\end{itemize}

This paper is organized as follows. \S\ref{sec:background} introduces cloud RDBMSs and analyzes current compression approaches in RDBMSs and their challenges. \S\ref{sec:evolution} presents our dual-layer compression design and database-oriented optimizations. \S\ref{sec:large} discusses challenges encountered in large-scale deployments and our corresponding solutions. Finally, \S\ref{sec:eval} comprehensively evaluates \sys's space utilization and performance, including an ablation study of individual techniques. 
\section{Background and Motivation}
\label{sec:background}
This section first introduces the architecture of \dbname, followed by a review of compression approaches in RDBMSs to reveal the motivation for developing \sys.

\subsection{The Architecture of \dbname}
\label{subsec:db}

The storage-compute separation architecture has been widely adopted by leading cloud-native RDBMSs, including AWS Aurora~\cite{DBLP:conf/sigmod/VerbitskiGSBGMK17}, Azure Hyperscale~\cite{DBLP:conf/sigmod/AntonopoulosBDS19}, and Alibaba PolarDB~\cite{feifeili, cloudjump, DBLP:journals/pvldb/CaoLWCZZWM18} as illustrated in \autoref{pic:polarFs}.
In this architecture, each database instance comprises a single read-write node (RW) responsible for handling both read and write requests, multiple read-only nodes (RO) dedicated to processing read-only queries, and a shared storage system.
For write operations, the RW node retrieves the necessary pages from the shared storage, generates redo logs, and subsequently transmits these logs to the corresponding storage nodes.
The storage nodes, which provide redundancy and high availability, ensure the durability of these redo logs and asynchronously apply them to generate updated pages.
To maintain transactional consistency, the RW node synchronizes transaction-related information in redo logs (i.e., the log record of transaction begins and transaction commits) to the RO nodes. Each RO node parses these redo logs to generate readviews for read transactions and maintains a local LSN ($\texttt{LSN}_{i}$), representing the progress of log parsing.
During read operations, RO nodes fetch required pages from storage nodes based on their $\texttt{LSN}_{i}$.
In the background, storage nodes track the minimum \texttt{LSN} across all RO nodes $\texttt{LSN}_{{\texttt{min}}}$ and apply redo logs up to this point. 
This design delegates page generation to storage nodes instead of having RW nodes generate and transmit pages, significantly reducing network bandwidth consumption while enabling instant startup of computing nodes during recovery or scaling operations.

Based on this architecture, cloud computing vendors provide flexible mechanisms of adjusting instance specifications (including CPU cores and memory of RW/RO nodes) based on dynamic use-case requirements, without any data migration overhead.
This capability significantly reduces the cost of computing resources for users.
However, storage costs remain a more important consideration for users, especially during periods of low activity.
Therefore, we introduce data compression in our RDBMSs to reduce storage costs.

\begin{figure}[]
  \begin{center}
    \includegraphics[width=\linewidth]{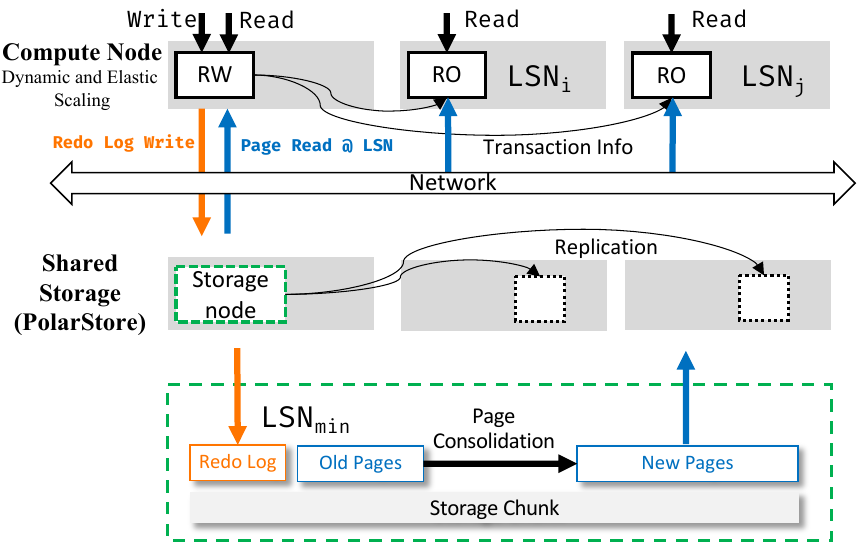}
  \end{center}
  \mycaption{pic:polarFs}{Architecture of \dbname}{}
\end{figure}

\subsection{Data Compression in RDBMSs}
Designing compression systems for RDBMSs requires careful consideration of the following two fundamental aspects.

\begin{figure} 
  \centering
  \subfloat[Index Granularity]{\includegraphics[trim=0 0 2.2in 0,clip,width=0.3715\linewidth]{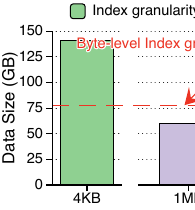}\label{fig:a}}
  \hfill
  \subfloat[Input Size]{\includegraphics[trim=1.3in 0 1.5in 0,clip,width=0.2\linewidth]{fig/moti2.pdf}\label{fig:b}}
  \hfill
  \subfloat[Algorithm]{\includegraphics[trim=2in 0 0 0,clip,width=0.4285\linewidth]{fig/moti2.pdf}\label{fig:c}}
  \mycaption{pic:moti}{Compressed storage sizes of a 408.37GB dataset under different configurations}{Including index granularity, input size, and algorithm. Red line: byte-level indexing, 16KB input size, and zstd, achieving 5.24$\times$ compression ratio.}
\end{figure}

\noindent\textbf{Index granularity.}
The size of compressed data varies according to the content, even when the original data size is fixed.
Therefore, systems must maintain an index to locate compressed data, and when the size changes (e.g., due to updates), allocate new storage space and update the index accordingly. 
This indexing mechanism presents a critical \textit{trade-off between space utilization and performance}. 
Large index granularity (e.g., 4\,KB-level) simplifies management but wastes storage space. 
For example, if a 16\,KB page compresses to 1\,KB, but the index granularity is 4\,KB, 3\,KB of space is wasted.
Our experiments with a 408.37\,GB dataset compressed using zstd show that 4KB index granularity consumes approximately 80.5\% more space than byte-level index granularity (Figure~\ref{fig:a}). 
However, while finer-grained indexing improves space efficiency, it introduces complex space management that degrades performance.

\noindent\textbf{Flexibility in compression parameters.}
The compression ratio is influenced by both the input block size and the chosen compression algorithm.
Larger blocks provide more opportunities to identify repeating patterns, thereby achieving better compression ratios. 
Our experiments demonstrate that 1\,MB blocks achieve a compression ratio of 6.85, while 4\,KB blocks achieves only 3.59 (Figure~\ref{fig:b}). 
Similarly, for compression algorithms, more advanced algorithms like Zstandard (\texttt{zstd}) consistently outperform simpler ones like \texttt{lz4} (Figure~\ref{fig:c}).
However, larger blocks and complex algorithms introduce performance overhead through I/O amplification and increased processing latency, respectively.
An effective system therefore needs to dynamically adjust these parameters based on access patterns: employing larger input sizes and complex algorithms for infrequently accessed cold data, while utilizing smaller compression input sizes with simpler algorithms for hot data and data on critical paths. 
However, this presents another \textit{trade-off between space utilization and performance}: while optimizing compression ratios, the flexibility in compression parameters introduces significant challenges for hardware acceleration implementation.

Current compression approaches in RDBMSs, whether software-based or hardware-based, fail to effectively balance these trade-offs between space utilization and performance.

\begin{table*}[]
  \resizebox{1\linewidth}{!}{
    \begin{tabular}{c|rlc}
      \toprule
      
      \textbf{Compression Approach}     & \textbf{Input Size $\Rightarrow$} & \textbf{Index Granularity} & 
      \textbf{Algorithm}  \\ \hline
      
      Compression in B+-Tree~\cite{tablecompr,pagecompr}    &  Flexible (\texttt{16KB} DB Page) $\Rightarrow$ & \cons{\texttt{4KB} File Blocks} &  Flexible  \\
      
      Compression in LSM-Tree~\cite{rocksDBcompr, levelcompr, yang2022oceanbase}  & Flexible (\texttt{16KB} DB Page) $\Rightarrow$ & \texttt{Bytes}, \cons{GC Overhead}  & Flexible \\
      
      Compression in Log-structured Block Storage~\cite{zhang2024s, li2023more} & Flexible (\texttt{16KB Segment}), \cons{I/O Amplification} $\Rightarrow$ & \texttt{Bytes}, \cons{GC Overhead}   &  Flexible  \\            
      \midrule
      In-Storage Compression~\cite{chen2024ha, DBLP:conf/fast/QiaoCZLL022, BreathingNewLife} & \cons{\texttt{4KB} LBA} $\Rightarrow$ & \texttt{Bytes}   & \cons{Inflexible}   \\ 
      Dedicated Compression Accelerators~\cite{mircosoft, chiosa2022hardware, qat}  & -  &   &  Flexible  \\   
      \midrule 
      \midrule 
      \sys & Flexible (\texttt{16KB} DB Page) $\Rightarrow$ \texttt{4KB} LBA $\Rightarrow$ & \texttt{Bytes}   & Flexible   \\ 
      \bottomrule
    \end{tabular} 
  }
  \mycaption{tbl:categories}{Comparison of data compression approaches in cloud RDBMSs}{Weaknesses highlighted in \cons{red}.}
\end{table*}

\subsubsection{Software-based Compression}
We analyze three software-based compression approaches implemented at different layers: B+-Tree and LSM-Tree at the database level, and log-structured storage at the storage level. 
Although these approaches offer flexibility in compression parameters, they all struggle to balance space efficiency with the resulting index management overhead.

\noindent\textbf{Compression in B+-Tree.}
Compression in B+-Trees (\myCircled{A} in \autoref{pic:vs}) is implemented through two distinct strategies.
The first strategy integrates compression directly into the tree structure by mapping each 16\,KB page to multiple 4\,KB blocks based on the compressed page size.
This approach handles updates by appending uncompressed data at the page's end and performs compression during page merge or split operations (e.g., InnoDB's table compression feature~\cite{tablecompr}).
The second strategy preserves the original tree structure while compressing pages prior to disk writes, utilizing file system hole-punching for space reclamation (e.g., InnoDB's page compression feature~\cite{pagecompr}).
However, B+-Trees suffer from inherent fragmentation, typically reserving approximately 20\% to 50\% of page space to accommodate future insertions~\cite{DBLP:conf/fast/QiaoCZLL022}.
Although this unused space can be compressed, the 4\,KB block granularity for indexing the compressed data still causes fragmentation and leads to suboptimal space efficiency.

\noindent
\textbf{Compression in LSM-Trees.}
LSM-Trees (\myCircled{B} in \autoref{pic:vs}), exemplified by RocksDB~\cite{rocksDBcompr}, LevelDB~\cite{levelcompr} and OceanBase~\cite{yang2022oceanbase}, integrate compression into their compaction mechanism.
During compaction operations, the data is compressed before being written to new SSTables, then the system updates its index to track the compressed data blocks.
While this approach achieves better space efficiency than B+-Trees by reducing fragmentation, it introduces substantial garbage collection overhead.
This overhead not only costs CPU resources but also competes with normal operations for I/O resources~\cite{hotstorageLSM,luo2020lsm,lu2013extending,matsunobu2020myrocks}.

\noindent
\textbf{Compression in log-structured block storage.}
Log-structured block storage systems (\myCircled{C} in \autoref{pic:vs}), such as Alibaba's Pangu~\cite{zhang2024s, li2023more}, employ compression during segment compaction.
While these systems encounter challenges similar to LSM-Trees, they suffer an additional performance penalty due to misalignment between compression units and database access units.
When a database page spans multiple compressed units, accessing a single page requires multiple read and decompression operations.

\begin{figure}[]
  \begin{center}
    \includegraphics[width=\linewidth]{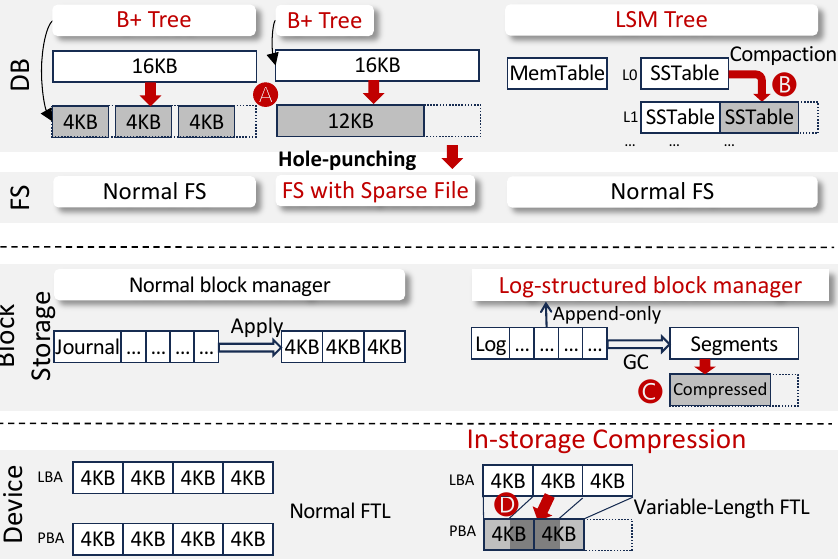}
  \end{center}
  \mycaption{pic:vs}{Data compression approaches in cloud RDBMSs}{The comparison of these approaches is shown in \autoref{tbl:categories}.} 
\end{figure}

\subsubsection{Hardware-based Compression}
Hardware-based approaches address performance overhead by offloading compression tasks to specialized accelerators, although this typically comes at the cost of reduced flexibility in compression input size and algorithm selection.

\noindent
\textbf{In-storage compression.}
Computational storage devices (CSDs, \myCircled{D} in \autoref{pic:vs}) integrate computational and memory resources within storage devices~\cite{lambdaIO, omnicache, CSDvirtualization, DBLP:conf/fast/CaoLCZLWOWWKLZZ20, chen2024ha, DBLP:conf/fast/QiaoCZLL022, BreathingNewLife}.
They can offload both compression/decompression tasks and index management to in-storage components, with the latter handled by the flash translation layer (FTL).
By extending the FTL to support non-aligned Physical Block Addresses (PBA) and leveraging existing garbage collection mechanisms, CSDs enable efficient, fine-grained indexing without software overhead.
However, their in-storage compression is constrained by fixed 4KB compression input sizes (for NVMe compatibility) and immutable compression algorithms set during manufacturing, limiting their flexibility for dynamic data characteristics and access patterns. 

\noindent \textbf{Dedicated compression accelerators.} 
Various hardware accelerators exist for compression/decompression, including PCIe-attached FPGAs/ASICs~\cite{mircosoft, chiosa2022hardware} and CPU-based accelerators such as Intel QAT~\cite{qat}. 
While they effectively reduce computational overhead, they do not address the fundamental challenge of index management. 

\noindent 
In summary, as illustrated in \autoref{tbl:categories}, existing RDBMS compression approaches face different challenges: software-based approaches offer flexibility in compression parameters but incur performance overhead for byte-level indexing, while hardware-based approaches improve performance but lack adaptability to varying workloads.
This limitation motivates us to develop a novel approach that combines the advantages of both software-based and hardware-based approaches.

\begin{figure*}[]
  \begin{center}
    \includegraphics[width=0.75\linewidth]{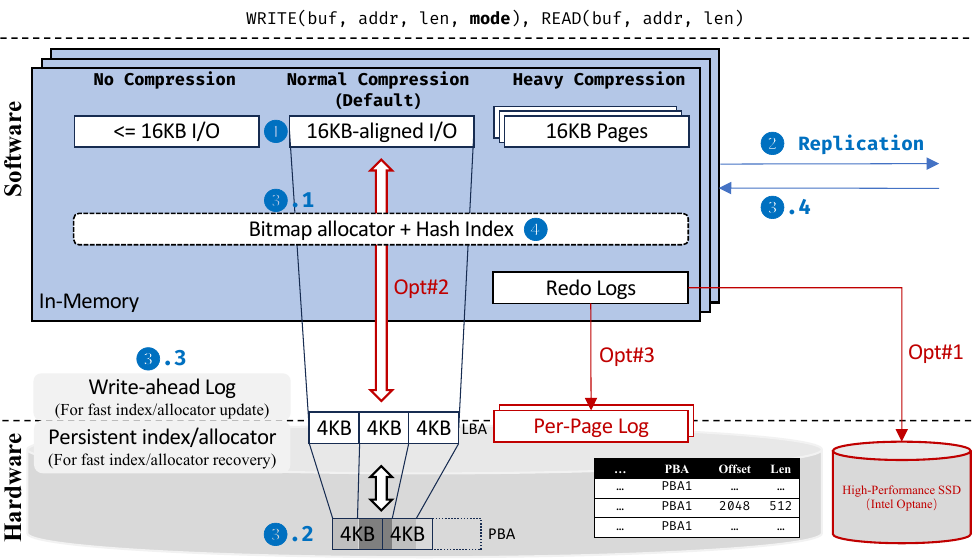}
  \end{center}
  \mycaption{pic:scs}{Overview of \sys}{There are two innovations: dual-layer compression with hardware-software co-designing and DB-oriented I/O optimizations (Opt). This figure also shows the workflow of a 16\,KB \texttt{write} with normal compression. }
\end{figure*}

\section{Design of \sys}

\subsection{Overview and Key Ideas}

In this paper, we propose \sys, a shared storage system for RDBMSs that achieves both high space utilization and high performance through two key ideas:

\label{sec:evolution}
\noindent\textbf{Dual-layer compression.}
\sys implements a novel dual-layer compression architecture that combines software compression with in-storage compression. 
The software layer compresses user data into 4\,KB-aligned blocks, which are subsequently compressed by \dev~into byte-aligned blocks. 
This design effectively addresses two fundamental challenges: 
\begin{itemize}
 [itemsep=0pt, parsep=0pt, labelsep=3pt, leftmargin=*, topsep=0pt, partopsep=0pt]
\item For the granularity of the index, \sys achieves byte-level indexing granularity by leveraging the FTL's existing garbage collection mechanisms.
This design eliminates additional space management overhead in software, as the software layer only needs to manage 4\,KB-aligned blocks.
\item For compression parameter flexibility, the software layer enables selection of compression input sizes and algorithms for different workloads.
\end{itemize}

\noindent
\textbf{DB-oriented optimizations.}
While compression operations inherently introduce computational latency to I/O operations, only two specific I/O operations directly impact user-perceived performance in RDBMSs: redo log writes and page reads.
Redo log writes are critical during transaction commits, where storage nodes must persist log records and synchronize with replicas to ensure durability and high availability.
Page reads become performance bottlenecks when compute nodes need to fetch pages absent from their buffer pools.
Guided by these insights, \sys implements targeted optimizations for these critical operations, prioritizing their performance even at the cost of reduced efficiency in other noncritical operations.
These optimizations include: 
utilizing high-performance devices to bypass compression for redo log writes (Opt\#1), implementing an adaptive algorithm selection for accelerating page reads (Opt\#2), and introducing a per-page log mechanism to reduce their read amplification (Opt\#3). 

The remainder of this section details its core components: the dual-layer compression mechanism in \S\ref{design1} and the three DB-oriented optimizations in \S\ref{design2}.


\subsection{Dual-layer Compression}
\label{design1}
We present \sys's architecture through three key components: software design (\S\ref{subsec:softcom}), hardware implementation (\S\ref{subsec:hardware}), and the flexible compression interface (\S\ref{subsec:flexible}).

\subsubsection{Lightweight Compression in Software} 
\label{subsec:softcom}
As illustrated in \autoref{pic:scs}, \sys's dual-layer compression architecture exposes a block interface to the upper layer, with write operations controlled by a compression mode flag.
To demonstrate the design, consider a 16\,KB write request workflow.
Upon receiving such a request, the storage node (acting as the leader) first compresses the data into multiple 4\,KB blocks (\ding{182}).
For fault tolerance, \sys implements 3-way Raft replication, where the leader forwards the compressed data to two replica nodes (\ding{183}).
The write request achieves commitment only after the data has been persisted on a majority of replicas.
To ensure durability, both leader and follower nodes execute three steps: allocate space for compressed blocks (\ding{184}.1), write them to their CSDs (\ding{184}.2), and record index updates in their write-ahead logs (\ding{184}.3). 
After receiving sufficient acknowledgments from followers (\ding{184}.4), the leader updates its in-memory index cache to make the changes visible and signals completion to the upper layer (\ding{185}).

\sys's space management architecture comprises two primary components: a space allocator and a hash table index.
The space allocation mechanism operates at two levels: a centralized allocator manages space at 128\,KB granularity for each storage device, while individual logical chunks employ bitmap allocators for fine-grained 4\,KB space management.  
\sys implements a hash table index to maintain mappings between uncompressed 16\,KB addresses and their corresponding compressed 4\,KB addresses.
While the global allocator persists its state through in-place updates, both the bitmap allocator and hash table index operate in memory. Their modifications are logged in the write-ahead log, which serves exclusively for recovery purposes.

\subsubsection{In-storage Compression: \dev} 
\label{subsec:hardware}

This part presents the design overview of \dev, focusing on key components essential to understand the dual-layer compression mechanism. 
In \S\ref{subsec:hw2}, we discuss insights derived from our first-generation implementation that guided the evolution of our new-generation architecture. 

\dev~exposes standard NVMe interfaces to the software layer and implements the gzip algorithm with compression level 5, which has been demonstrated to achieve optimal performance in hardware acceleration scenarios~\cite{mircosoft}.
The space management system of \dev~extends the traditional page-mapping FTL architecture.
While conventional page-mapping FTL maintains fixed-length (4\,KB) mappings between Logical Block Addresses (LBA) and Physical Block Addresses (PBA), \dev~introduces variable-length index entries that support mappings from 4KB-aligned LBA to byte-level PBA.
In our implementation, \dev~augments each mapping entry with 12-bit \textit{length} and \textit{offset} fields to specify compressed data positions within a 4\,KB boundary. This enhancement adds 3\,bytes to each index entry, increasing the memory footprint from the original 5\,bytes (for the basic information of L2P mapping) to 8\,bytes per entry.

To maintain compatibility, we set \dev~with a logical capacity of 7.68\,TB, aligned with the standard capacity of mainstream enterprise SSDs.
The physical NAND flash capacity is dimensioned based on compression ratios in our target workloads. Our comprehensive evaluation of the gzip algorithm (compression level 5), configured to process 4\,KB-aligned inputs with byte-level output granularity, demonstrates an average compression ratio of 2.4 across diverse datasets.
Based on this compression efficiency, \dev~is provisioned with at least 3.2\,TB of physical NAND flash space to support the logical capacity of 7.68\,TB and the 2.4 compression ratio.

\subsubsection{Flexibility of Interface} 

\noindent
\textbf{Write interface.}
The storage layer extends its write interface with a flag field that supports three compression modes: normal compression, no compression, and heavy compression.

\noindent
\underline{Normal compression:} Following the workflow presented in \S\ref{subsec:softcom}, this default mode mandates that I/O operations be aligned to database page boundaries (typically 16\,KB or 8\,KB). Any non-aligned operation automatically reverts to the no-compression mode.

\noindent
\underline{No compression:} This mode serves two specific scenarios: handling non-page-aligned I/O operations and processing user-designated uncompressed pages.
The software bypasses compression and writes the data directly to \dev.
When writing to a part of a previously compressed range, \sys must decompress the existing data, allocate new storage space, and write the uncompressed data to \dev.

\noindent
\underline{Heavy compression:} This mode is specifically designed to re-store and compress a range of data (i.e., archiving operations). Unlike other modes, this interface does not write new data to the storage.
Instead, it processes the entire write range as a single compression unit and uses high-level compression configuration to achieve higher compression ratios.
It first reads and decompresses any existing compressed pages within the range, then merges them into a single segment and recompresses this consolidated segment for optimal compression ratios.
The compressed segment is stored contiguously, with each index entry in the hash table maintaining both the address of the compressed segment and the page offset within that segment.
While this approach may introduce I/O amplification during random access, such overhead is negligible for archived and snapshot pages in databases, which are typically accessed sequentially during analytical processing and backup/recovery operations, and are infrequently accessed. A temporary buffer for decompressed segments effectively optimizes such sequential access patterns.

\noindent
\textbf{Read interface.}
Unlike the write interface, the read interface does not require additional parameters.
The system maintains three essential attributes in each index entry: compression status, compression algorithm used, and segment information for heavily compressed pages. These attributes guide the system in determining the appropriate read range and decompression strategy during read operations.

\label{subsec:flexible}

\subsection{DB-oriented Optimizations}
\label{design2}

Building upon the architectural overview, we introduce three DB-oriented optimizations that minimize compression overhead by exploiting database-specific access patterns, with a focus on reducing the latency of redo writes and page reads. These optimizations include (\autoref{pic:scs}):
\begin{itemize}
[itemsep=0pt, parsep=0pt, labelsep=3pt, leftmargin=*, topsep=0pt, partopsep=0pt]
\item (Opt\#1) Utilizing high-performance devices to bypass compression for latency-sensitive redo log writes (\S\ref{subsec:tech1})
\item (Opt\#2) Implementing adaptive algorithm selection between lz4 and zstd to balance compression ratios and read latency (\S\ref{subsec:tech2})
\item (Opt\#3) Introducing a per-page log mechanism that leverages physical-logical space decoupling to reduce read amplification during page accesses (\S\ref{subsec:tech3})
\end{itemize}

\subsubsection{Avoiding Compression for \textit{Log Write}}
\label{subsec:tech1}
Redo log persistence is critical for transaction commit latency.
To minimize this latency, \sys bypasses all compression for redo logs: it employs the no-compression flag to skip software compression and leverages high-performance storage to avoid hardware compression.
In our implementation, each server incorporates an Intel Optane SSD, which delivers superior and stable performance with limited space.
Initially, these devices were dedicated to storing WAL for in-memory data structures (the allocator and index), and their usage has now been extended to include redo logs as well. The use of Optane SSDs for redo logs is a good match for their characteristics: the limited capacity of these devices is sufficient
for redo logs, which are typically small and can be reclaimed after pages are flushed.


\begin{figure}
  \centering
  \subfloat[Decompr Lat ($\mu s$)]{\includegraphics[trim=0 0 2in 0,clip,width=0.428\linewidth]{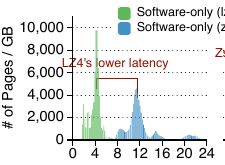}\label{lz4:a}} 
  \hfill
  \subfloat[Algo-level Ratio]{\includegraphics[trim=1.5in 0 1in 0,clip,width=0.2855\linewidth]{fig/lightv0.pdf}\label{lz4:b}} 
  \hfill
  \subfloat[Dual-layer Ratio]{\includegraphics[trim=2.5in 0 0 0,clip,width=0.2855\linewidth]{fig/lightv0.pdf}\label{lz4:c}} 
  \mycaption{pic:lightv0}{Performance and compression ratio analysis of LZ4 and Zstd}{(a) Decompression latency confirms zstd's higher computational cost. (b) 
  At the software level, zstd offers a superior compression ratio. (c) However, in our dual-layer system, zstd's advantage is substantially diminished.}
\end{figure}

\subsubsection{Low-latency I/O and Decompression for \textit{Page Read}}
\label{subsec:tech2}
The I/O latency of page reads directly impacts query latency when requested pages are absent from the in-memory buffer pool. These page reads include two parts: storage read latency and decompression latency. To reduce these latencies, we introduce an adaptive algorithm selection mechanism.
Our adaptive algorithm selection mechanism is motivated by two key observations. 

First, conventional wisdom holds that zstd, despite its longer decompression times (Figure~\ref{lz4:a}), achieves superior compression ratios compared to lz4 (Figure~\ref{lz4:b}).
In the algorithm layer, there is a simple trade-off between performance and compression ratios.
However, as shown in Figure~\ref{lz4:c}, our comprehensive evaluation reveals an interesting phenomenon in the dual-layer compression setting: its compression advantage significantly diminishes from 58.9\% at the algorithm level to merely 9.0\% after hardware compression. 
This substantial reduction occurs because hardware gzip, which also employs Huffman encoding, can effectively further compress lz4's output (which lacks Huffman encoding) while gaining minimal additional compression on zstd's output (which already incorporates Huffman encoding). This observation reveals that: 
\textit{for some pages, using lz4 offers both lower decompression latency and a competitive compression ratio compared to using zstd.}

Second, the 4\,KB I/O alignment of compressed pages creates another powerful optimization opportunity. Even a marginal size reduction at the software layer from zstd can be enough to save an entire 4\,KB I/O block. For instance, if lz4 compresses a 16\,KB page to 4097\,bytes (requiring 8\,KB of I/O in data write/read) while zstd achieves 4096\,bytes (requiring only 4\,KB of I/O), the I/O savings in data read can easily outweigh zstd's higher decompression overhead. This observation demonstrates a counter-intuitive scenario where:
\textit{for certain pages, using zstd can achieve a lower total page read latency than lz4.}

Taken together, these observations challenge the notion of a simple trade-off between latency and compression ratio in using lz4 or zstd. Instead, they reveal the potential for a "\textit{win-win}" outcome: the compression algorithm is no longer a static choice between two competing algorithms, but a dynamic decision to pick the clear winner for each specific page. 
Therefore, \sys implements a page-level algorithm selection mechanism during page writes, as shown in Algorithm~\ref{algo}. It evaluates both algorithms' 4KB ceiling-aligned compressed sizes and decompression latencies (\texttt{Line 6-8}). If the ratio of saved data size (i.e., the benefits of zstd) to increased decompression latency (i.e., the overhead of zstd) exceeds a threshold (\texttt{Line 15}), \sys switches to zstd for this page; otherwise, it stays with lz4 to maintain low decompression overhead.

\begin{algorithm}
  \caption{Selection of lz4 and zstd}
  \label{algo}
  \SetKwFunction{lzx}{\textbf{lz4}}
  \SetKwFunction{zstdx}{\textbf{zstd}}
  \SetKwFunction{lastx}{\textbf{last\_used\_algorithm}}
  \SetKwFunction {myfunc}{page\_compression}
  \SetKwProg{Fn}{Function}{:}{}
  \Fn{\myfunc{\texttt{page}}}{
     \If{\texttt{CPU\_utilization} > 20\%} {
      \texttt{ptr $\leftarrow$} \lzx{\texttt{}}\\
      \textbf{return} \texttt{"lz4", ptr}
      }
    \ElseIf{\texttt{update\_percent} > 30\%} {
      // \texttt{lz4\_sz}, \texttt{zstd\_sz}: 4KB ceiling-aligned\\
      \texttt{lz4\_ptr, lz4\_sz, lz4\_lat}$\leftarrow$\lzx{}\\
      \texttt{zstd\_ptr, zstd\_sz, zstd\_lat}$\leftarrow$\zstdx{}\\
      \,\\
        // extra decompression latency of zstd: $\mu s$\\
        \texttt{overhead $\leftarrow$ zstd\_lat - lz4\_lat}\\
        // saved storage space of zstd: bytes\\
        \texttt{benefit $\leftarrow$ lz4\_sz - zstd\_sz}\\
        \,\\
        \If{\texttt{benefit / overhead > 300B/$\mu s$}} {
            \textbf{return} "zstd"\texttt{, zstd\_ptr}
        }
        \Else{
            \textbf{return} "lz4"\texttt{, lz4\_ptr}
        }
    }
    \Else{
        \texttt{ptr $\leftarrow$} \lastx{\texttt{}}\\
        \textbf{return} \texttt{last\_used\_algorithm, ptr}
    }
  }
\end{algorithm}

This threshold is set to 300\,B/$\mu s$ based on storage I/O characteristics: saving 4\,KB of I/O typically reduces read latency by 12\textasciitilde14\,$\mu s$, translating to approximately 300\,B/$\mu s$. When zstd's storage savings per additional microsecond of decompression time exceed this threshold, the I/O latency reduction outweighs the increased decompression overhead, making zstd the better choice.

It is important to note that this selection occurs during page writes, which are out of the critical path of user queries. However, the process still consumes CPU resources, which must be carefully managed to avoid impacting overall system throughput. 
To minimize the selection overhead, this mechanism is triggered only in two cases: during initial page writes, or when the database layer estimates page updates exceed 30\% based on the log size. Additionally, the selection process only runs when CPU utilization is low, ensuring that this selection does not become a performance bottleneck.

\subsubsection{Mitigating the Tail Latency for \textit{Page Read}}
\label{subsec:tech3}

\label{sec:encoding}

\label{subsec:algorithm}


\label{subsec:pagelog}

We leverage CSD's space decoupling feature to address the tail latency issue in page read operations.

\begin{figure} 
  \centering
  \subfloat[Traditional Page Consolidation]{\includegraphics[trim=0 0 9.03cm 0,clip,width=0.5\linewidth]{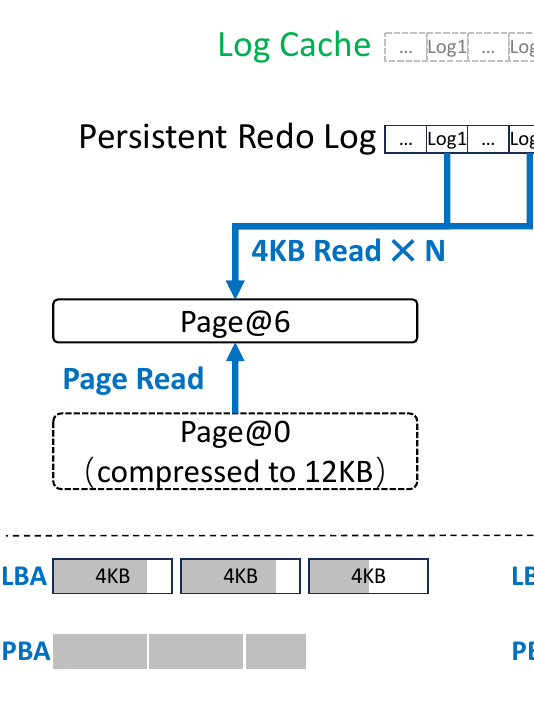}\label{pagea}}
  \hfill
  \subfloat[Per-page Log Optimization]{\includegraphics[trim=9.03cm 0 0 0,clip,width=0.5\linewidth]{pic/pageserver.pdf}\label{pageb}}
  \mycaption{pic:perpage}{Workflow of page consolidation}{(a) Traditional method: scattered redo logs lead to read amplification and high tail-latency. (b) Per-page log optimization: proactively co-locating logs in the background enables page consolidation with a single I/O.}
\end{figure}

For page reads, both the database buffer pool and storage software memory cache help avoid storage accesses and decompression overhead. However, when pages are not cached in memory, I/O becomes unavoidable. Three scenarios may occur: (i) the page exists but is not cached, (ii) the page does not exist but its redo logs are cached, and (iii) the page does not exist and some of its redo logs are not cached. The third case commonly occurs when a RO node's LSN falls behind due to high load or network issues, preventing the storage node from recycling redo logs and leading to cache overflow.
While the first two cases require only a single page read, the third case suffers from read amplification, requiring multiple random reads to retrieve redo logs. As illustrated in Figure~\autoref{pagea}, to generate \texttt{page@6}, the storage node must read both $log_1$ and $log_3$ from storage, which may reside in different 4KB address ranges. These scattered reads significantly contribute to slow tail latencies. 

\noindent
\textbf{Design with space decoupling.}
CSD decouples logical space allocation from physical space utilization, allowing software to manage logical space at 4\,KB granularity while utilizing physical space more flexibly. This enables implementing sparse data structures without I/O amplification or space waste~\cite{DBLP:conf/fast/QiaoCZLL022,BreathingNewLife}.
\sys introduces a per-page log mechanism. The key is co-locating all redo logs of each page within a dedicated 4\,KB log space when the log is evicted from the in-memory cache. As shown in Figure~\autoref{pageb}, when the redo log is evicted from memory (this occurs before receiving a read request for \texttt{page@6}), the storage node pre-merges both $log_1$ and $log_3$ into the per-page log space. This allows retrieving all necessary logs in a single read, when the storage node receives the read request for \texttt{page@6}, significantly reducing read amplification and improving tail latency.
This solution, allocating an additional 4\,KB log space for each 16\,KB page, is only feasible with CSD's space decoupling feature. Implementing such a design on conventional SSDs would incur approximately 25\% space amplification.

\section{Large-Scale Deployment of \sys} 
\label{sec:large}

We initially deployed our system across 11 clusters with 500 hosts, containing a total of 6000 \dev~devices. Through this large-scale deployment, we identified two critical challenges: host-level stability and cluster-level resource utilization. In \S\ref{subsec:hw2}, we address the host-level stability challenge by summarizing lessons learned from the first generation hardware (\dev1.0) and redesigning a new generation device (\dev2.0). In \S\ref{subsec:cluster}, we tackle the cluster-level resource utilization challenge by introducing a compression-aware scheduling technique that effectively balances chunks with varying compression ratios across the cluster.

\subsection{Host-Level Stability}
\label{subsec:hw2}

The first generation of \dev~was designed with an open-channel architecture (host-based FTL)~\cite{lu2013extending, lu2019ocstore,zhang2017flashkv,oclu1,oclu2}, which offered advantages in rapid development and flexible space management. While we initially dedicated specific CPU cores to FTL tasks to minimize its impact on software performance, our long-term deployment of 12 devices per host revealed significant challenges in system stability and operational costs.

\subsubsection{Lessons in Host-level Stability} 
During an eighteen-month period, there were 26 occurrences of slow I/O (i.e., exceeding 1 second) caused by \dev1.0. Among these, 6 slow I/O events exceeded 10 seconds in latency and lasted for more than 10 minutes, adversely affecting the user's business and significantly increasing the maintenance overhead. These problems arise from two factors: resource contention and expanded fault domain for host-based FTL driver.

\noindent
\textbf{Resource contention for host-based FTL.}
The host-based FTL consumes host memory and CPU resources. This overhead becomes more severe when supporting variable-length address mapping and multiple drives. We observed 12 occurrences of slow I/O caused by memory contention and 9 occurrences caused by CPU contention. Each \dev1.0, with a logical capacity of 7.68TB, requires 15.36\,GB ($7.68TB \times 8B / 4KB$) of memory for FTL. With 12 devices per storage node, the total memory consumption reaches approximately 184.32\,GB. This significant memory footprint causes resource pressure, triggering aggressive memory reclamation by the system that degrades application performance. Further, the host-based FTL for each \dev1.0 requires approximately 2 dedicated physical CPU cores to maintain performance under high-pressure workloads. Consequently, each server needs 24 physical CPU cores, and these FTL threads may interfere with the original storage software, leading to I/O jitters.

\noindent
\textbf{Expanded fault domain.}
As a new product, the open-channel driver of \dev~contained some undiscovered bugs. When triggered by a single device, they could affect the entire server. Our observation showed that all 5 long-lasting slow I/O occurrences were caused by such kernel driver bugs.

As a temporary solution to these problems, we had to disable software compression in our dual-layer compression design and limit the deployment to 10 devices per server, ensuring sufficient resources for the host-based FTL. However, this compromise reduced our storage density due to fewer devices per server (from 12 to 10) and the loss of software compression benefits.

\subsubsection{Solution: \dev2.0}
Based on the lessons from \dev1.0, we design a new generation CSD (\dev2.0) that addresses the above stability issues through several key improvements:

First, \dev2.0 abandons the open-channel architecture and reverts to conventional device-managed FTL, where the embedded ARM cores handle LBA-to-PBA (L2P) mapping and background operations. This return to traditional architecture eliminates the resource contention issues of host-based FTL and contains FTL failures within individual devices, preventing them from affecting the entire host system.
Second, to improve storage density, we increase the NAND flash capacity in \dev2.0 to 3.84\,TB (using 4\,TB NAND flash with 4\% over-provisioning), enabling 20\% more compressed data storage per device. 
Third, to maintain the overall compression ratio of 2.4 with this increased storage capacity, we optimize the FTL mapping structure to avoid additional memory consumption. To reduce FTL memory consumption, we redesigned the L2P mapping entry in \dev2.0. While \dev1.0 used an 8-byte entry, \dev2.0 reduces this to 7\,bytes. This memory saving is achieved by coarsening the physical offset granularity from a single byte to 16 bytes, which allows the required offset and length metadata to be encoded in 2 bytes instead of 3. This optimization effectively addresses the memory requirements from the increased storage capacity and allows us to expose 9.6\,TB of logical space for each \dev2.0 device. 
Finally, following the industry trend, \dev2.0 adopts PCIe 4.0 to enhance I/O performance.

\subsubsection{The Evaluation of \dev2.0}
\label{subsub:2.0}
\begin{figure}[]
  \begin{center}
    \includegraphics[width=\linewidth]{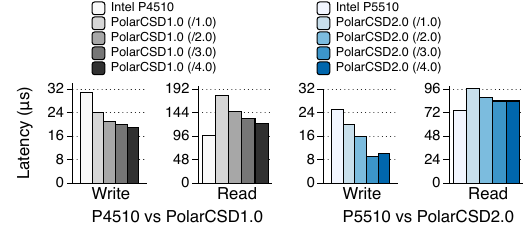}
  \end{center}
  \mycaption{pic:ssd1}{Average latency of \dev~and standard SSDs under different compression ratios}{Workload: 16KB I/O, queue-depth=1. Target compression ratios: 1.0, 2.0, 3.0, 4.0 (configured via FIO~\cite{fio}).}
\end{figure}

\begin{figure}[]
  \begin{center}
    \includegraphics[width=\linewidth]{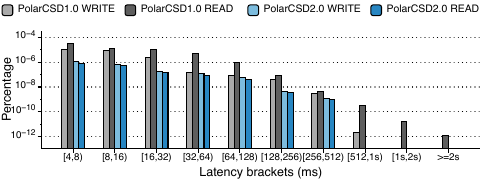}
  \end{center}
  \mycaption{fig:tail}{Distribution of device latency (>=4ms) in production with 4\textasciitilde16KB READ and WRITE operations}{}
\end{figure}

We first conduct basic performance testing. For fair comparison, we use Intel P4510 (PCIe~3.0) and P5510 (PCIe~4.0) as the baseline for \dev1.0 (PCIe~3.0) and \dev2.0 (PCIe~4.0) respectively, matching their respective PCIe interfaces. \autoref{pic:ssd1} presents the latency under workloads with different compression ratios. The results show that \dev1.0 achieves lower write latency but higher read latency compared to P4510. We also observe that larger compression ratios lead to lower latencies, as less data needs to be written to or read from NAND flash.

After large-scale deployment in our production environment, we observe significant improvements in system stability. First, there is no slow I/O caused by CPU/memory contention, and the failure of a single device does not impact the whole host.
Further, the tail latency of \dev2.0 is also significantly reduced. 
\autoref{fig:tail} shows the latency distribution of these two generations of PolarCSD in our production environment during 7 days. We only show the percentage of I/Os with latency larger than 4ms in this figure, and we observe that only $7.91\times 10^{-7}$ read and $1.05\times 10^{-6}$ write latency of \dev2.0 are larger than 4ms.  In contrast, \dev1.0 shows much higher percentages ($2.9\times 10^{-5}$ and $4.0\times 10^{-5}$, respectively), approximately 36.7 times and 38.8 times higher than \dev2.0.

\begin{figure}[]
    \centering
    \subfloat[ ]{%
        \includegraphics[width=0.5\linewidth]{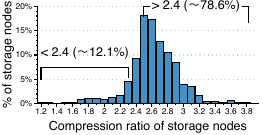}%
        \label{fig:distribution}%
    }
    \hfill
    \subfloat[ ]{%
        \includegraphics[width=0.5\linewidth]{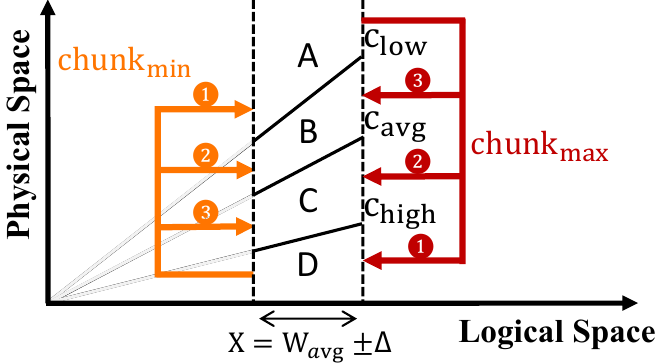}%
        \label{fig:ScheProcess}%
    }
    \mycaption{fig:sched}{(a) Distribution of compression ratio in a cluster, and (b) Compression-aware scheduling algorithm}{}
\end{figure}
\subsection{Cluster-level Space Management}
\label{subsec:cluster}
During our large-scale deployment, we discovered that the original data scheduling strategy fails to effectively handle compressed data, leading to suboptimal resource utilization.

\subsubsection{Lessons in Cluster-level Resource Utilization} 
Initially, our clusters employed a simple scheduling strategy based solely on logical space usage: new chunks were allocated to storage nodes with the lowest logical space usage. When a node's logical space usage exceeded the average usage $w_{avg}$ by 10\%, chunks would be migrated to nodes with the lowest logical space usage. Nodes exceeding 75\% space usage (any of logical space usage or physical space usage) were blocked from receiving new chunks, and when all nodes reached this threshold, manual intervention was required to add new storage servers.
During deployment, we identified two major limitations in our original space management:

\noindent
\textbf{Inaccurate physical space monitoring.} 
The software can query devices for physical space usage, but the lack of TRIM operations leads to inaccurate measurements. When our software allocator frees space, it only updates space management metadata without actually releasing physical space via TRIM operations, and therefore devices remain unaware of these releases, causing reported physical space usage to exceed actual usage. After enabling TRIM operations upon space deallocation, the monitored physical space decreased by 3\% on average.

\noindent
\textbf{Compression ratio imbalance.} 
The original strategy failed to account for varying compression ratios across different chunks, leading to significant space waste. When the cluster became full, we observed an interesting phenomenon: some storage nodes reached their 75\% logical space limit while their physical spaces remained underutilized, while other nodes showed the opposite pattern. Simply redistributing chunks between these two types of nodes could have increased the cluster's effective capacity without adding new servers. Our analysis of a full cluster running \dev1.0 revealed the extent of this imbalance (Figure~\ref{fig:distribution}): 12.1\% of storage nodes had below-average compression ratios, wasting 1.72\% of total logical space, while 78.6\% had above-average ratios, wasting 9.17\% of total physical space.
\begin{figure}[]
    \centering
    \subfloat[Before Scheduling.]{%
        \includegraphics[width=0.48\linewidth]{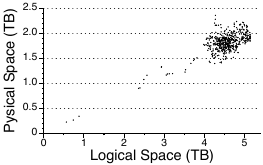}%
        \label{fig:C1DisBefore}%
    }
    \hfill
    \subfloat[After Scheduling.]{%
        \includegraphics[width=0.48\linewidth]{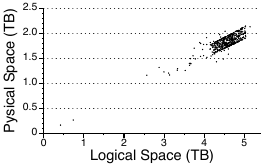}%
        \label{fig:C1DisAfter}%
    }
    \mycaption{fig:before}{Logical-to-physical space mapping of storage instances before and after scheduling}{Clusters with hardware-only compression (\dev1.0).}
\end{figure}

\begin{figure}[]
    \centering
    \subfloat[Before Scheduling.]{%
        \includegraphics[width=0.48\linewidth]{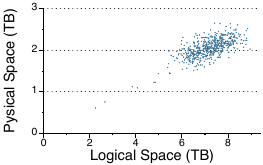}%
        \label{fig:C2DisBefore}%
    }
    \hfill
    \subfloat[After Scheduling.]{%
        \includegraphics[width=0.48\linewidth]{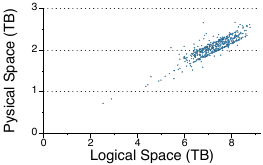}%
        \label{fig:C2DisAfter}%
    }
    \mycaption{fig:c2}{Logical-to-physical space mapping of storage instances before and after scheduling}{Clusters with dual-layer compression (\dev2.0 and software compression).}
    \label{}
\end{figure}


\subsubsection{Solution: Compression-aware Scheduling}
To balance both logical and physical space usage, we developed a compression-aware scheduling strategy. As shown in Figure~\ref{fig:ScheProcess}, storage nodes can be visualized on a two-dimensional plane with logical space (x-axis) and physical space (y-axis). Since the cluster allocates new chunks to nodes with the lowest logical space usage, most nodes are distributed between $x=w_{avg}\pm\Delta$, where $w_{avg}$ is the average logical space usage.
Our strategy aims to maintain compression ratios within a range $[c_l , c_h]$, where $c_l < c_{avg} < c_h$ and $c_{avg}$ is the average compression ratio. These values divide the operational region into four zones: Zone A (high physical, low logical usage), Zone B (balanced usage, below average), Zone C (balanced usage, above average), and Zone D (low physical, high logical usage). For nodes in Zone A, \sys migrates chunks with minimum compression ratios to nodes in zones D, C, or B (in order of preference). Conversely, for nodes in Zone D, chunks with maximum compression ratios are migrated to nodes in zones A, B, or C.

\subsubsection{Scheduling Results in Production Clusters}

The selection of $c_l$ and $c_h$ is a trade-off between the scheduling results and the number of scheduling tasks. Generally, lower $c_l$ and higher $c_h$ result in fewer tasks. We determine these parameters through offline simulations for each cluster, targeting the parameters completion within one day. 
We evaluated our strategy on two production clusters: C1 (using \dev~1.0 without software compression) and C2 (using \dev~2.0 with software compression). Figure~\ref{fig:C1DisBefore} and Figure~\ref{fig:C1DisAfter} show the distribution of logical and physical spaces in C1 before and after scheduling, while Figure~\ref{fig:C2DisBefore} and Figure~\ref{fig:C2DisAfter} show results for C2. 
After scheduling, storage nodes converged into a distinct quadrilateral region, with over 90\% of nodes in C1 achieving compression ratios between 2.2 and 2.7, and 87.7\% of nodes in C2 between 3.15 and 3.85, demonstrating effective compression ratio balance.

\begin{figure*}[]
  \centering
  \subfloat[SQL: Throughput]{%
      \includegraphics[trim=0 0 4.8in 0,clip,width=0.3333\textwidth]{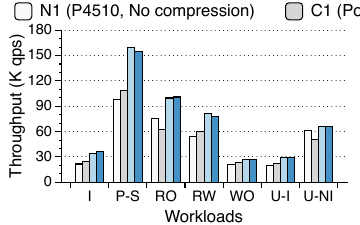}%
      \label{fig:all1}%
  }
  \hfill
  \subfloat[SQL:  Average Latency]{%
  \includegraphics[trim=2.4in 0 2.4in 0,clip,width=0.3333\textwidth]{fig/all.pdf}%
      \label{fig:all2}%
  }
  \hfill
  \subfloat[SQL: P95 Latency]{%
  \includegraphics[trim=4.8in 0 0 0,clip,width=0.3333\textwidth]{fig/all.pdf}%
      \label{fig:all3}%
  }
  \mycaption{pic:all}{Overall performance}{Configurations: Sysbench~\cite{sysbench}, 16 threads in a single client. Workloads: \textbf{I}: Insert, \textbf{P-S}: Point Select, \textbf{RO}: OLTP-Read-Only, \textbf{RW}: OLTP-Read-Write, \textbf{WO}: OLTP-Write-Only, \textbf{U-I}: Update-Index, \textbf{U-NI}: Update-Non-Index.}
\end{figure*}

\section{Evaluation}
\label{sec:eval}

In this section, we seek to answer the following questions:
\begin{itemize}
  [itemsep=0pt, parsep=0pt, labelsep=3pt, leftmargin=*, topsep=0pt, partopsep=0pt]
\item What are the space savings and performance benefits of \sys in large-scale production deployments (\S\ref{eval1})?
\item How much does each technique contribute to \sys's overall effectiveness (\S\ref{subsec:ablation})?
\item How does \sys compare with existing software compression approaches (\S\ref{other})?
\end{itemize}


\subsection{Cluster-level Evaluation}
\label{eval1}
Our production environment consists of over 500 storage servers with more than 6000 first-generation computational storage devices (\dev1.0) and over 1200 storage servers with over 14400 second-generation computational storage devices (\dev2.0), respectively. We maintain two types of clusters: The first type, represented by cluster C1, uses \dev1.0, where we had to disable software compression and two DB-oriented optimizations (i.e., the selection mechanism of lz4/zstd and the per-page log mechanism) to mitigate resource contention issues. The second type, represented by cluster C2, uses \dev2.0~with all techniques enabled. 
To evaluate the effectiveness of these two compressed clusters, we compare their space utilization and database performance with two normal clusters (N1 and N2) selected from our production environment. These normal clusters use contemporary Intel SSDs (P4510 in N1, P5510 in N2) and have identical hardware specifications (CPU, PCIe, and NICs) to C1 and C2 respectively. Detailed configurations are shown in \autoref{tbl:ratio}.

\subsubsection{Space Utilization and Cost Analysis}
The cluster C1, equipped with \dev1.0, achieves a 2.35 compression ratio in production. Although the hardware cost of \dev1.0 is \texttt{1.45} higher than the baseline Intel P4510 (normalized as \texttt{1.00}) due to additional embedded memory and accelerators, the effective cost per GB of logical storage reduces to \texttt{0.62} after compression.
The cluster C2, equipped with \dev2.0 and dual-layer compression, demonstrates even better results. Through hardware optimization, \dev2.0's relative cost reduces by 9\% compared to \dev1.0 (from \texttt{1.45} to \texttt{1.32}). More importantly, with the addition of software compression, C2 achieves a significantly higher compression ratio of 3.55. This brings the effective cost per GB down to 0.37, representing about 60\% reduction in storage cost compared to using Intel P5510 (0.91).

\begin{table}[]
  \resizebox{1\linewidth}{!}{
    \begin{tabular}{c|cc|cc}
      \toprule      
      \textbf{Cluster}& \textbf{N1}  &  \textbf{C1}  & \textbf{N2}  & \textbf{C2} \\   
      \midrule
      \textbf{Software}& -    & -   & -    & \textbf{Compression} \\   
      \textbf{Hardware}& P4510    & \textbf{\dev1.0}   & P5510    &  \textbf{\dev2.0} \\   
      \textbf{Opt\#1: bypass redo}& -    & \ding{52}   & -    & \ding{52} \\  
      \textbf{Opt\#2: lz4/zstd}& -    & \ding{56}   & -    & \ding{52} \\  
      \textbf{Opt\#3: per-page log}& -    & \ding{56}   & -    & \ding{52} \\  
      
      \textbf{Scheduling}& -    & \ding{52}   & -    & \ding{52} \\  
      \midrule
      \textbf{NAND Size}& 3.84~TB     &    3.20~TB     &    7.68~TB  & 3.84~TB  \\      

      \textbf{Compression Ratio}& -   &      2.35    & - &   3.55 \\   
      \textbf{Cost/GB(Physical)}& 1   &      1.45    & 0.91 &   1.32 \\
      \textbf{Cost/GB(Logical)}& 1   &      0.62    & 0.91 &   0.37 \\
      
      \midrule
      \textbf{CPU}&  \multicolumn{2}{c|}{Xeon Platinum 2.5GHz}  &     \multicolumn{2}{c}{Xeon Platinum 2.9GHz}  \\  
      \textbf{NIC} & \multicolumn{2}{c|}{CX-4 (25Gbps$\times$2)} & \multicolumn{2}{c}{CX-6 (100Gbps$\times$2)} \\
      \textbf{PCIe} & \multicolumn{2}{c|}{3.0} & \multicolumn{2}{c}{4.0} \\
      
      \textbf{\# of storage devices}& 12      &    10   &      12  & 12 \\  
      \textbf{Performance layer} & \multicolumn{2}{c|}{Intel P4800X} & \multicolumn{2}{c}{Intel P5800X} \\    
      
         
      \bottomrule
    \end{tabular} 
  }
  \mycaption{tbl:ratio}{Cluster configurations and space utilization}{}
\end{table}

\begin{figure*}[]
  \centering
  \subfloat[SQL: Throughput]{%
      \includegraphics[trim=0 0 5.76in 0,clip,width=0.2\textwidth]{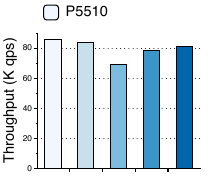}%
      \label{fig:1}%
  }
  \hfill
  \subfloat[SQL: Avg Latency]{%
  \includegraphics[trim=1.44in 0 4.32in 0,clip,width=0.2\textwidth]{fig/final.pdf}%
      \label{fig:2}%
  }
  \hfill
  \subfloat[I/O: Redo Write Latency]{%
  \includegraphics[trim=2.88in 0 2.88in 0,clip,width=0.2\textwidth]{fig/final.pdf}%
      \label{fig:3}%
  }
  \hfill
  \subfloat[I/O: Page Read Latency]{%
  \includegraphics[trim=4.32in 0 1.44in 0,clip,width=0.2\textwidth]{fig/final.pdf}%
      \label{fig:4}%
  }
  \hfill
  \subfloat[I/O: Page Write Latency]{%
  \includegraphics[trim=5.76in 0 0 0,clip,width=0.2\textwidth]{fig/final.pdf}%
      \label{fig:5}%
  }
  \mycaption{fig:all}{Impact of techniques on performance}{This figure contrasts user-level performance with internal I/O latency: (a) and (b) show key user-request metrics, while (c), (d), and (e) detail the components of I/O latency at the storage layer.}
\end{figure*}

\subsubsection{Performance Evaluation}
We use Sysbench~\cite{sysbench} to evaluate database performance. The database compute instance runs with 8 cores and 32\,GB memory, accessing data distributed across 8 storage nodes (48 chunks in total). The database size is configured to 480\,GB with 60\,GB data per storage node (120\,GB with replication). Database requests are generated by a separate Elastic Compute Service (ECS) client using 16 threads. The limited memory configuration creates an I/O-bound environment, ideal for evaluating storage system compression/decompression performance. 
We evaluate throughput (Figure~\ref{fig:all1}), average latency (Figure~\ref{fig:all2}), and P95 latency (Figure~\ref{fig:all3}) across different workloads. While C1 (\dev1.0) shows a 10\% performance degradation compared to N1 (P4510), our latest C2 cluster (\dev2.0) achieves performance parity with N2 (P5510), demonstrating the effectiveness of our hardware and software optimizations.

\subsection{Ablation Study of Techniques}
\label{subsec:ablation}
To quantify the contribution of each technique, we conduct ablation studies on both performance and space utilization. We run the baseline on N2 cluster with P5510 SSDs, while on C2 cluster we add optimization technique one at a time to evaluate its impact. For performance evaluation (\autoref{fig:all}), we use the identical benchmark environments as described above. All experiments use Sysbench OLTP-Read-Write workload. For space utilization (\autoref{pic:softratio}), we use N2's storage size as baseline and measure the relative compression ratios achieved in C2. We evaluate compression effectiveness using four production datasets that were dumped from user databases and restored in our experimental databases with permission.

\noindent
\textbf{Base hardware compression \texttt{(\dev)}.} With only \dev's hardware compression enabled, we observe compression ratios ranging from 2.12$\times$ to 3.84$\times$ across different datasets, but at the cost of 7.4\% throughput reduction compared to P5510 (Figure~\ref{fig:1}). This performance degradation stems from \dev's higher read latency affecting page read operations (Figure~\ref{fig:4}).

\noindent
\textbf{Dual-layer compression \texttt{(+dual-layer)}.}
Adding software compression (using zstd by default) further improves compression ratios by 21.7\%\textasciitilde50.3\% but leads to a 19.6\% throughput reduction compared to using hardware compression alone. This performance drop mainly occurs because compressing 16KB redo writes in the software layer slows down redo write operations, increasing their average latency (with 3-way replication and durability) from 59$\mu s$ to 79$\mu s$, as shown in Figure~\ref{fig:3}.

\begin{figure}[]
  \begin{center}
    \includegraphics[width=\linewidth]{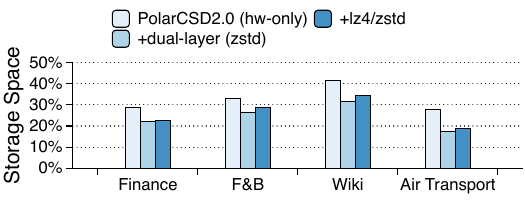}
  \end{center}
  \vspace{-1em}
  \mycaption{pic:softratio}{Impact of techniques on space utilization}{}
\end{figure}


\noindent
\textbf{Avoiding compression for log write \texttt{(+bypass redo)}.} By bypassing compression for redo writes, we reduce the throughput degradation to only 8.9\% compared to hardware compression. Since redo logs are small and frequently recycled, this optimization maintains the compression ratio for user data.

\noindent
\textbf{Low-latency I/O and decompression for page read via compression selection \texttt{(+lz4/zstd)}.} 
When we enable the algorithm selection between lz4 and zstd, system throughput improves significantly, approaching the baseline performance (only 2.1\% lower as shown in Figure~\ref{fig:1}). This improvement stems from our adaptive strategy that optimally balances I/O and decompression overhead: choosing lz4 when faster decompression is beneficial, and selecting zstd when its additional compression can reduce I/O operations. This optimization reduces average page read latency by approximately 9$\mu s$ compared to using zstd exclusively. While this selection mechanism increases page write latency, as shown in Figure~\ref{fig:5}, this overhead occurs in the background and does not directly impact user operations. 
Note that, in our evaluation, the update is always issue the algorithm re-selection, representing the worstpage write latency. And in real workloads, algorithm re-selection is unfrenquent, which does not bring high write latency.   
The impact on compression ratio is minimal. As shown in \autoref{pic:softratio}, using compression selection only increases storage space by 0.7\%\textasciitilde2.6\% compared to using zstd exclusively. The distribution of pages compressed by zstd versus lz4 varies across different datasets, as shown in \autoref{tbl:rate}. 

\begin{table}[]
  \centering
  \resizebox{0.8\linewidth}{!}{
    \centering
    \begin{tabular}{c|cccc}
      \toprule 
      Dataset & \textbf{Finance} & \textbf{F\&B}  &  \textbf{Wiki}  & \textbf{Air Transport}  \\  
      \midrule
      zstd &  73.1\% &  41.3\% & 52.4\%  & 51.6\%  \\  
      lz4 & 26.9\% &  58.7\%  & 47.5\% &   48.4\% \\  
      \bottomrule
    \end{tabular} 
  }
  \mycaption{tbl:rate}{Distribution of selected compression algorithms (zstd vs. lz4) across different database workloads}{}
\end{table}


\begin{figure}
  \centering
  \subfloat[Throughput]{\includegraphics[trim=0 0 2.8in 0,clip,width=0.333\linewidth]{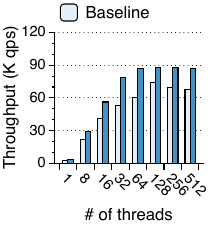}\label{perpage:a}} 
  \hfill
  \subfloat[Average Latency]{\includegraphics[trim=1.4in 0 1.4in 0,clip,width=0.333\linewidth]{fig/perpage.pdf}\label{perpage:b}} 
  \hfill
  \subfloat[P95 Latency]{\includegraphics[trim=2.8in 0 0 0,clip,width=0.333\linewidth]{fig/perpage.pdf}\label{perpage:c}} 
  \mycaption{pic:perpagelog}{Performance of OLTP read-only (RO nodes) before and after using per-page log optimizations}{}
  \vspace{-0.1in}
\end{figure}

\noindent
\textbf{Reducing the tail latency for page read via per-page log.}
To evaluate the effectiveness of our per-page log design under insufficient log cache conditions, we set up a two-node experiment: a read-write (RW) node and a read-only (RO) node, with the RO node intentionally lagging approximately 1s behind in LSN synchronization. This setup prevents log recycling at storage nodes and ensures log cache pressure. We direct OLTP write-only workloads to the RW node and read-only workloads to the RO node. 
As shown in \autoref{pic:perpagelog}, we measure performance on the RO node with varying thread counts. With threads below 128, the per-page log optimization significantly reduces P95 latency by 28.9\%\textasciitilde39.5\% compared to the baseline. This improvement stems from reduced I/O amplification during page generation, as our design enables retrieving all necessary logs with a single read operation instead of multiple scattered reads. However, beyond 128 threads, the performance becomes CPU-bound at the RO node, where software queuing dominates the latency and diminishes the benefits of our I/O optimization.




\begin{figure}
  \centering
  \subfloat[Throughput]{\includegraphics[trim=0 0 2.4in 0,clip,width=0.333\linewidth]{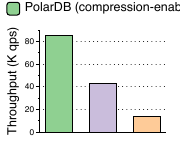}\label{apple:a}} 
  \hfill
  \subfloat[Average Latency]{\includegraphics[trim=1.2in 0 1.2in 0,clip,width=0.333\linewidth]{fig/apple.pdf}\label{apple:b}} 
  \hfill
  \subfloat[P95 Latency]{\includegraphics[trim=2.4in 0 0 0,clip,width=0.333\linewidth]{fig/apple.pdf}\label{apple:c}} 
  \mycaption{pic:apple}{End-to-end performance comparison with other approaches}{}
  \vspace{-0.1in}
\end{figure}

\subsection{Comparison with Other Approaches}
\label{other}
We compare \dbname~that enables compression~\cite{polarstore} with two other popular databases that implement compression at the database layer: InnoDB with table compression~\cite{tablecompr} and MyRocks with compression enabled~\cite{matsunobu2020myrocks}.
\autoref{pic:apple} shows throughput and latency comparison using Sysbench OLTP-Read-Write workload. This experiment uses the same configurations as previous ones. \dbname~demonstrates superior performance over other systems. This is because InnoDB and MyRocks consume resources in compute nodes (which are billed to users) for both space management and compression/decompression operations. These approaches force users to make a trade-off between storage and computing resources, which does not truly reduce costs.
In contrast, \sys implements compression at the shared storage layer, which offers two key advantages: first, compression becomes completely transparent to database users without consuming their compute resources. Second, by centrally managing resources used for compression across multiple users, cloud providers can potentially achieve better resource utilization and lower operational costs.








\section{Discussion}

\label{discussion}

This section discusses related research directions on compression in RDBMSs and alternative space-saving approaches.

\noindent
\textbf{Related Directions on Data Compression.}
There are also some solutions can further improve data compression in RDBMSs.
First, to improve compression ratios, we can leverage table-level information to generate shared dictionaries~\cite{dict-kempa2018roots,dict-larsson2000off} for pages within the same table. This approach improves compression ratios by exploiting schema-level semantics and reducing per-page metadata overhead.
Second, we can optimize page layouts to facilitate dictionary construction. Database systems store structured data with well-defined field boundaries and type information. By leveraging these properties and incorporating specialized data encoding techniques, we can achieve substantially better compression ratios compared to general-purpose algorithms. Additionally, clustering data of the same column together~\cite{slesarev2021revisiting,DBLP:conf/sigmod/AbadiMF06,DBLP:conf/vldb/PossP03} can enhance compression effectiveness by exploiting column-specific patterns.
Third, estimation techniques~\cite{harnik2013zip} can enable rapid algorithm selection to better balance performance and compression ratios.
Fourth, leveraging hardware acceleration and vector instructions available in modern CPUs~\cite{qat,abali2020data} can further improve compression/decompression performance.

\noindent
\textbf{Alternative Space-Saving Approaches.}
Storage tiering~\cite{tiered-wang2020bcw, tiered-zheng2019ziggurat,guerra2011cost, zhang2010adaptive} offers another approach to cost reduction. For instance, our system supports table-level archiving of cold data to object storage.
Erasure coding (EC)~\cite{ec-lee2016erasure,ec-li2013erasure,ec-li2019openec,ec-plank2005t1} presents an alternative for reducing storage costs while maintaining reliability comparable to replication-based systems. However, EC is not currently suitable for our system's redo records and remains a future research direction.
Further, data deduplication~\cite{harnik2016estimating, dedup-fast,dedup-xia2016fastcdc,dedup-chen2016ordermergededup,dedup-fu2015design,dedup-yang2019smartdedup, hotstorage, dedup-cao2019sliding,dedup-cao2023smrts} can effectively reduce storage costs, but its applicability in RDBMSs is limited since data is typically stored at the record level, making exact page-level deduplication matches rare.



\section{Conclusion}
This paper revisits data compression in RDBMS and presents PolarStore, a shared storage system with dual-layer compression architecture for large-scale cloud-native databases. The key contributions include a hardware/software co-designed compression mechanism that achieves both high space utilization and flexibility, database-oriented optimizations that maintain high performance on critical I/O paths, and practical solutions for large-scale deployment challenges. PolarStore has been deployed on thousands of nodes, serving tens of thousands of customers, and represents one of the largest known deployments of computational storage in production databases.

\section*{Acknowledgements}
We sincerely thank our shepherd Juncheng Yang and anonymous reviewers for their valuable feedback, which significantly improved this paper. We thank the PolarDB Infrastructure Team and Alibaba Infrastructure Service (AIS) Group for their support and contributions to this work. 

\bibliographystyle{unsrt}
\bibliography{sample.bib}

@article{DBLP:journals/pvldb/CaoLWCZZWM18,
  author    = {Wei Cao and Zhenjun Liu and Peng Wang and Sen Chen and Caifeng Zhu and Song Zheng and Yuhui Wang and Guoqing Ma},
  title     = {{PolarFS}: An Ultra-low Latency and Failure Resilient Distributed File System for Shared Storage Cloud Database},
  journal   = {Proc. {VLDB} Endow.},
  volume    = {11},
  number    = {12},
  pages     = {1849--1862},
  year      = {2018},
  month     = {aug},
  doi       = {10.14778/3229863.3229872},
  url       = {https://doi.org/10.14778/3229863.3229872}
}

@article{matsunobu2020myrocks,
  author    = {Yoshinori Matsunobu and Siying Dong and Herman Lee},
  title     = {{MyRocks}: {LSM-Tree} Database Storage Engine Serving Facebook's Social Graph},
  journal   = {Proc. {VLDB} Endow.},
  volume    = {13},
  number    = {12},
  pages     = {3217--3230},
  year      = {2020},
  publisher = {{VLDB Endowment}}
}

@article{huang2020tidb,
  author    = {Dongxu Huang and Qi Liu and Qiu Cui and Zhuhe Fang and Xiaoyu Ma and Fei Xu and Li Shen and Liu Tang and Yuxing Zhou and Menglong Huang and others},
  title     = {{TiDB}: A {Raft-based} {HTAP} Database},
  journal   = {Proc. {VLDB} Endow.},
  volume    = {13},
  number    = {12},
  pages     = {3072--3084},
  year      = {2020},
  publisher = {{VLDB Endowment}}
}

@inproceedings{DBLP:conf/vldb/PossP03,
  author    = {Meikel P{\"{o}}ss and Dmitry Potapov},
  title     = {Data Compression in Oracle},
  journal   = {Proc. {VLDB} Endow.},
  pages     = {937--947},
  year      = {2003},
  publisher = {Morgan Kaufmann},
  doi       = {10.1016/B978-012722442-8/50087-2},
  url       = {http://www.vldb.org/conf/2003/papers/S28P01.pdf}
}

@inproceedings{DBLP:conf/sigmod/AbadiMF06,
  author    = {Daniel J. Abadi and Samuel Madden and Miguel Ferreira},
  title     = {Integrating Compression and Execution in {Column-Oriented} Database Systems},
  booktitle = {ACM {SIGMOD} International Conference on Management of Data (SIGMOD '06)},
  pages     = {671--682},
  year      = {2006},
  publisher = {{ACM}},
  doi       = {10.1145/1142473.1142548},
  url       = {https://doi.org/10.1145/1142473.1142548}
}

@inproceedings{DBLP:conf/hotstorage/ZhengCLWLPSZ020,
  author    = {Ning Zheng and Xubin Chen and Jiangpeng Li and Qi Wu and Yang Liu and Yong Peng and Fei Sun and Hao Zhong and Tong Zhang},
  title     = {{Re-think} Data Management Software Design Upon the Arrival of Storage Hardware with {Built-in} Transparent Compression},
  booktitle = {12th {USENIX} Workshop on Hot Topics in Storage and File Systems (HotStorage '20)},
  year      = {2020},
  publisher = {{USENIX} Association},
  url       = {https://www.usenix.org/conference/hotstorage20/presentation/zheng}
}

@inproceedings{DBLP:conf/fast/QiaoCZLL022,
  author    = {Yifan Qiao and Xubin Chen and Ning Zheng and Jiangpeng Li and Yang Liu and Tong Zhang},
  title     = {Closing the {B+-tree} vs. {LSM-tree} Write Amplification Gap on Modern Storage Hardware with {Built-in} Transparent Compression},
  booktitle = {20th {USENIX} Conference on File and Storage Technologies (FAST '22)},
  pages     = {69--82},
  year      = {2022},
  publisher = {{USENIX} Association},
  url       = {https://www.usenix.org/conference/fast22/presentation/qiao}
}

@article{yang2022oceanbase,
  title={OceanBase: a 707 million tpmC distributed relational database system},
  author={Yang, Zhenkun and Yang, Chuanhui and Han, Fusheng and Zhuang, Mingqiang and Yang, Bing and Yang, Zhifeng and Cheng, Xiaojun and Zhao, Yuzhong and Shi, Wenhui and Xi, Huafeng and others},
  journal={Proceedings of the VLDB Endowment},
  volume={15},
  number={12},
  pages={3385--3397},
  year={2022},
  publisher={VLDB Endowment}
}

@article{BreathingNewLife,
  author    = {Kecheng Huang and Zhaoyan Shen and Zili Shao and Tong Zhang and Feng Chen},
  title     = {Breathing New Life into an Old Tree: Resolving Logging Dilemma of {B+-tree} on Modern Computational Storage Drives},
  journal   = {Proc. {VLDB} Endow.},
  volume    = {17},
  number    = {2},
  pages     = {134--147},
  year      = {2023},
  month     = {oct},
  doi       = {10.14778/3626292.3626297},
  url       = {https://doi.org/10.14778/3626292.3626297},
  issn      = {2150-8097},
  publisher = {{VLDB Endowment}}
}

@inproceedings{DBLP:conf/sigmod/VerbitskiGSBGMK17,
  author    = {Alexandre Verbitski and Anurag Gupta and Debanjan Saha and Murali Brahmadesam and Kamal Gupta and Raman Mittal and Sailesh Krishnamurthy and Sandor Maurice and Tengiz Kharatishvili and Xiaofeng Bao},
  title     = {Amazon Aurora: Design Considerations for High Throughput {Cloud-Native} Relational Databases},
  booktitle = {{ACM} International Conference on Management of Data (SIGMOD '17)},
  pages     = {1041--1052},
  year      = {2017},
  publisher = {{ACM}},
  doi       = {10.1145/3035918.3056101},
  url       = {https://doi.org/10.1145/3035918.3056101}
}

@inproceedings{DBLP:conf/sigmod/AntonopoulosBDS19,
  author    = {Panagiotis Antonopoulos and Alex Budovski and Cristian Diaconu and Alejandro Hernandez Saenz and Jack Hu and Hanuma Kodavalla and Donald Kossmann and Sandeep Lingam and Umar Farooq Minhas and Naveen Prakash and Vijendra Purohit and Hugh Qu and Chaitanya Sreenivas Ravella and Krystyna Reisteter and Sheetal Shrotri and Dixin Tang and Vikram Wakade},
  title     = {Socrates: The New {SQL} Server in the Cloud},
  booktitle = {International Conference on Management of Data (SIGMOD '19)},
  pages     = {1743--1756},
  year      = {2019},
  publisher = {{ACM}},
  doi       = {10.1145/3299869.3314047},
  url       = {https://doi.org/10.1145/3299869.3314047}
}

@article{feifeili,
  author    = {Feifei Li},
  title     = {Cloud Native Database Systems at Alibaba: Opportunities and Challenges},
  journal   = {Proc. {VLDB} Endow.},
  volume    = {12},
  number    = {12},
  pages     = {2263--2272},
  year      = {2019},
  doi       = {10.14778/3352063.3352141},
  url       = {http://www.vldb.org/pvldb/vol12/p2263-li.pdf}
}

@misc{polarstore,
  title        = {Enable the Storage Compression Feature in {PolarDB}},
  howpublished = {[Online]},
  note         = {\url{https://www.alibabacloud.com/help/en/polardb/polardb-for-mysql/how-to-turn-on-storage-compression}}
}

@misc{lz4,
  title        = {{LZ4}},
  howpublished = {[Online]},
  note         = {\url{https://github.com/lz4/}}
}

@misc{zstd,
  title        = {Zstandard ({ZSTD})},
  howpublished = {[Online]},
  note         = {\url{https://github.com/facebook/zstd}}
}

@misc{tablecompr,
  title        = {{InnoDB} Table Compression},
  howpublished = {[Online]},
  note         = {\url{https://dev.mysql.com/doc/refman/8.4/en/innodb-table-compression.html}}
}

@misc{sysbench,
  title        = {{Sysbench}},
  howpublished = {[Online]},
  note         = {\url{https://github.com/akopytov/sysbench}}
}

@misc{qat,
  title        = {{Intel QAT}},
  howpublished = {[Online]},
  note         = {\url{https://www.intel.com/content/www/us/en/architecture-and-technology/intel-quick-assist-technology-overview.html}}
}

@misc{pagecompr,
  title        = {{InnoDB} Page Compression},
  howpublished = {[Online]},
  note         = {\url{https://dev.mysql.com/doc/refman/8.4/en/innodb-page-compression.html}}
}

@misc{rocksDBcompr,
  title        = {Compression in {RocksDB}},
  howpublished = {[Online]},
  note         = {\url{https://github.com/facebook/rocksdb/wiki/Compression}}
}

@misc{levelcompr,
  title        = {Compression in {LevelDB}},
  howpublished = {[Online]},
  note         = {\url{https://github.com/google/leveldb/blob/main/doc/index.md}}
}

@inproceedings{zhang2024s,
  author    = {Weidong Zhang and Erci Xu and Qiuping Wang and Xiaolu Zhang and Yuesheng Gu and Zhenwei Lu and Tao Ouyang and Guanqun Dai and Wenwen Peng and Zhe Xu and others},
  title     = {What's the Story in {EBS} Glory: Evolutions and Lessons in Building Cloud Block Store},
  booktitle = {22nd {USENIX} Conference on File and Storage Technologies (FAST '24)},
  pages     = {277--291},
  year      = {2024}
}

@inproceedings{li2023more,
  author    = {Qiang Li and Qiao Xiang and Yuxin Wang and Haohao Song and Ridi Wen and Wenhui Yao and Yuanyuan Dong and Shuqi Zhao and Shuo Huang and Zhaosheng Zhu and others},
  title     = {More Than Capacity: {Performance-oriented} Evolution of Pangu in Alibaba},
  booktitle = {21st {USENIX} Conference on File and Storage Technologies (FAST '23)},
  pages     = {331--346},
  year      = {2023}
}

@inproceedings{lambdaIO,
  author    = {Zhe Yang and Youyou Lu and Xiaojian Liao and Youmin Chen and Junru Li and Siyu He and Jiwu Shu},
  title     = {{{\(\lambda\)}-IO}: {A} Unified {IO} Stack for Computational Storage},
  booktitle = {21st {USENIX} Conference on File and Storage Technologies (FAST '23)},
  pages     = {347--362},
  year      = {2023},
  publisher = {{USENIX} Association},
  url       = {https://www.usenix.org/conference/fast23/presentation/yang-zhe}
}

@inproceedings{CSDvirtualization,
  author    = {Dongup Kwon and Dongryeong Kim and Junehyuk Boo and Wonsik Lee and Jangwoo Kim},
  title     = {A Fast and Flexible {Hardware-based} Virtualization Mechanism for Computational Storage Devices},
  booktitle = {{USENIX} Annual Technical Conference (USENIX ATC '21)},
  pages     = {729--743},
  year      = {2021},
  publisher = {{USENIX} Association},
  url       = {https://www.usenix.org/conference/atc21/presentation/kwon}
}

@inproceedings{DBLP:conf/fast/CaoLCZLWOWWKLZZ20,
  author    = {Wei Cao and Yang Liu and Zhushi Cheng and Ning Zheng and Wei Li and Wenjie Wu and Linqiang Ouyang and Peng Wang and Yijing Wang and Ray Kuan and Zhenjun Liu and Feng Zhu and Tong Zhang},
  title     = {{{POLARDB}} Meets Computational Storage: Efficiently Support Analytical Workloads in {Cloud-Native} Relational Database},
  booktitle = {18th {USENIX} Conference on File and Storage Technologies (FAST '20)},
  pages     = {29--41},
  year      = {2020},
  publisher = {{USENIX} Association},
  url       = {https://www.usenix.org/conference/fast20/presentation/cao-wei}
}

@inproceedings{TerseCades,
  author    = {Gennady Pekhimenko and Chuanxiong Guo and Myeongjae Jeon and Peng Huang and Lidong Zhou},
  title     = {{TerseCades}: Efficient Data Compression in Stream Processing},
  booktitle = {{USENIX} Annual Technical Conference (USENIX ATC '18)},
  year      = {2018},
  pages     = {307--320},
  month     = {jul},
  publisher = {{USENIX} Association},
  url       = {https://www.usenix.org/conference/atc18/presentation/pekhimenko}
}

@inproceedings{cockshott1998data,
  author    = {W. Paul Cockshott and D. McGregor and Nikolaos Kotsis and John Wilson},
  title     = {Data Compression in Database Systems},
  booktitle = {Ninth International Workshop on Database and Expert Systems Applications (DEXA '98)},
  pages     = {981--990},
  year      = {1998},
  publisher = {IEEE}
}

@inproceedings{iyer1994data,
  author    = {Balakrishna R. Iyer and David Wilhite},
  title     = {Data Compression Support in Databases},
  journal   = {Proc. {VLDB} Endow.},
  pages     = {695--704},
  year      = {1994}
}

@article{gao2024revisiting,
  author    = {Chuqing Gao and Shreya Ballijepalli and Jianguo Wang},
  title     = {Revisiting {B-tree} Compression: An Experimental Study},
  journal   = {Proceedings of the ACM on Management of Data},
  volume    = {2},
  number    = {3},
  pages     = {1--25},
  year      = {2024},
  publisher = {{ACM}}
}

@inproceedings{chen2024ha,
  author    = {Xiang Chen and Tao Lu and Jiapin Wang and Yu Zhong and Guangchun Xie and Xueming Cao and Yuanpeng Ma and Bing Si and Feng Ding and Ying Yang and others},
  title     = {{HA-CSD}: Host and {SSD} Coordinated Compression for Capacity and Performance},
  booktitle = {{IEEE} International Parallel and Distributed Processing Symposium (IPDPS '24)},
  pages     = {825--838},
  year      = {2024},
  publisher = {IEEE}
}

@inproceedings{slesarev2021revisiting,
  author    = {Alexander Slesarev and Evgeniy Klyuchikov and Kirill Smirnov and George Chernishev},
  title     = {Revisiting Data Compression in {Column-Stores}},
  booktitle = {10th International Conference on Model and Data Engineering (MEDI '21)},
  pages     = {279--292},
  year      = {2021},
  publisher = {Springer}
}

@inproceedings{QATFS,
  author    = {Xiaokang Hu and Fuzong Wang and Weigang Li and Jian Li and Haibing Guan},
  title     = {{QZFS}: {QAT} Accelerated Compression in File System for Application Agnostic and Cost Efficient Data Storage},
  booktitle = {{USENIX} Annual Technical Conference (USENIX ATC '19)},
  year      = {2019},
  pages     = {163--176},
  month     = {jul},
  publisher = {{USENIX} Association},
  url       = {https://www.usenix.org/conference/atc19/presentation/hu-xiaokang}
}

@inproceedings{compressimage,
  author    = {Daniel Reiter Horn and Ken Elkabany and Chris Lesniewski-Lass and Keith Winstein},
  title     = {The Design, Implementation, and Deployment of a System to Transparently Compress Hundreds of Petabytes of Image Files for a {File-Storage} Service},
  booktitle = {14th {USENIX} Symposium on Networked Systems Design and Implementation (NSDI '17)},
  year      = {2017},
  pages     = {1--15},
  month     = {mar},
  publisher = {{USENIX} Association},
  url       = {https://www.usenix.org/conference/nsdi17/technical-sessions/presentation/horn}
}

@inproceedings{mao2017elastic,
  author    = {Bo Mao and Hong Jiang and Suzhen Wu and Yaodong Yang and Zaifa Xi},
  title     = {Elastic Data Compression with Improved Performance and Space Efficiency for {Flash-Based} Storage Systems},
  booktitle = {{IEEE} International Parallel and Distributed Processing Symposium (IPDPS '17)},
  pages     = {1109--1118},
  year      = {2017},
  publisher = {IEEE}
}

@article{chiosa2022hardware,
  author    = {Monica Chiosa and Fabio Maschi and Ingo M{\"{u}}ller and Gustavo Alonso and Norman May},
  title     = {Hardware Acceleration of Compression and Encryption in {SAP} {HANA}},
  journal   = {Proc. {VLDB} Endow.},
  volume    = {15},
  number    = {12},
  pages     = {3277--3291},
  year      = {2022},
  publisher = {{ACM}}
}

@inproceedings{omnicache,
  author    = {Jian Zhang and Yujie Ren and Marie Nguyen and Changwoo Min and Sudarsun Kannan},
  title     = {{OmniCache}: Collaborative Caching for {Near-storage} Accelerators},
  booktitle = {22nd {USENIX} Conference on File and Storage Technologies (FAST '24)},
  year      = {2024},
  pages     = {35--50},
  month     = {feb},
  publisher = {{USENIX} Association},
  url       = {https://www.usenix.org/conference/fast24/presentation/zhang-jian}
}

@article{cloudjump,
  author    = {Zongzhi Chen and Xinjun Yang and Feifei Li and Xuntao Cheng and Qingda Hu and Zheyu Miao and Rongbiao Xie and Xiaofei Wu and Kang Wang and Zhao Song and Haiqing Sun and Zechao Zhuang and Yuming Yang and Jie Xu and Liang Yin and Wenchao Zhou and Sheng Wang},
  title     = {{CloudJump}: Optimizing Cloud Databases for Cloud Storages},
  journal   = {Proc. {VLDB} Endow.},
  volume    = {15},
  number    = {12},
  pages     = {3432--3444},
  year      = {2022},
  month     = {aug},
  doi       = {10.14778/3554821.3554834},
  url       = {https://doi.org/10.14778/3554821.3554834},
  publisher = {{VLDB Endowment}}
}

@inproceedings{mircosoft,
  author    = {Jeremy Fowers and Joo-Young Kim and Doug Burger and Scott Hauck},
  title     = {A Scalable {High-Bandwidth} Architecture for Lossless Compression on {FPGAs}},
  booktitle = {{IEEE} 23rd Annual International Symposium on {Field-Programmable} Custom Computing Machines (FCCM '15)},
  year      = {2015},
  pages     = {52--59},
  doi       = {10.1109/FCCM.2015.46}
}

@inproceedings{hotstorageLSM,
  author    = {Gunhee Choi and Kwanghee Lee and Myunghoon Oh and Jongmoo Choi and Jhuyeong Jhin and Yongseok Oh},
  title     = {A New {LSM-style} Garbage Collection Scheme for {ZNS} {SSDs}},
  booktitle = {12th {USENIX} Workshop on Hot Topics in Storage and File Systems (HotStorage '20)},
  year      = {2020},
  month     = {jul},
  publisher = {{USENIX} Association},
  url       = {https://www.usenix.org/conference/hotstorage20/presentation/choi}
}

@article{luo2020lsm,
  author    = {Chen Luo and Michael J. Carey},
  title     = {{LSM-based} Storage Techniques: A Survey},
  journal   = {The VLDB Journal},
  volume    = {29},
  number    = {1},
  pages     = {393--418},
  year      = {2020},
  publisher = {Springer}
}

@inproceedings{dedup-cao2019sliding,
  author    = {Zhichao Cao and Shiyong Liu and Fenggang Wu and Guohua Wang and Bingzhe Li and David H. C. Du},
  title     = {Sliding {Look-Back} Window Assisted Data Chunk Rewriting for Improving Deduplication Restore Performance},
  booktitle = {17th {USENIX} Conference on File and Storage Technologies (FAST '19)},
  pages     = {129--142},
  year      = {2019}
}

@inproceedings{dedup-cao2023smrts,
  author    = {Zhichao Cao and Hao Wen and Fenggang Wu and David H. C. Du},
  title     = {{SMRTS}: A Performance and {Cost-Effectiveness} Optimized {SSD-SMR} Tiered File System with Data Deduplication},
  booktitle = {2023 {IEEE} 41st International Conference on Computer Design (ICCD '23)},
  pages     = {275--282},
  year      = {2023},
  publisher = {IEEE}
}

@inproceedings{dedup-xia2016fastcdc,
  author    = {Wen Xia and Yukun Zhou and Hong Jiang and Dan Feng and Yu Hua and Yuchong Hu and Qing Liu and Yucheng Zhang},
  title     = {{FastCDC}: A Fast and Efficient {Content-Defined} Chunking Approach for Data Deduplication},
  booktitle = {{USENIX} Annual Technical Conference (USENIX ATC '16)},
  pages     = {101--114},
  year      = {2016}
}

@inproceedings{dedup-fu2015design,
  author    = {Min Fu and Dan Feng and Yu Hua and Xubin He and Zuoning Chen and Wen Xia and Yucheng Zhang and Yujuan Tan},
  title     = {Design Tradeoffs for Data Deduplication Performance in Backup Workloads},
  booktitle = {13th {USENIX} Conference on File and Storage Technologies (FAST '15)},
  pages     = {331--344},
  year      = {2015}
}

@inproceedings{dedup-chen2016ordermergededup,
  author    = {Zhuan Chen and Kai Shen},
  title     = {{OrderMergeDedup}: Efficient, {Failure-Consistent} Deduplication on Flash},
  booktitle = {14th {USENIX} Conference on File and Storage Technologies (FAST '16)},
  pages     = {291--299},
  year      = {2016}
}

@inproceedings{dedup-yang2019smartdedup,
  author    = {Qirui Yang and Runyu Jin and Ming Zhao},
  title     = {{SmartDedup}: Optimizing Deduplication for {Resource-Constrained} Devices},
  booktitle = {{USENIX} Annual Technical Conference (USENIX ATC '19)},
  pages     = {633--646},
  year      = {2019}
}

@inproceedings{tiered-wang2020bcw,
  author    = {Shucheng Wang and Ziyi Lu and Qiang Cao and Hong Jiang and Jie Yao and Yuanyuan Dong and Puyuan Yang},
  title     = {{BCW}: {Buffer-Controlled} Writes to {HDDs} for {SSD-HDD} Hybrid Storage Server},
  booktitle = {18th {USENIX} Conference on File and Storage Technologies (FAST '20)},
  pages     = {253--266},
  year      = {2020}
}

@inproceedings{tiered-zheng2019ziggurat,
  author    = {Shengan Zheng and Morteza Hoseinzadeh and Steven Swanson},
  title     = {Ziggurat: A Tiered File System for {Non-Volatile} Main Memories and Disks},
  booktitle = {17th {USENIX} Conference on File and Storage Technologies (FAST '19)},
  pages     = {207--219},
  year      = {2019}
}

@inproceedings{dedup-fast,
  author    = {Kiran Srinivasan and Timothy Bisson and Garth R. Goodson and Kaladhar Voruganti},
  title     = {{iDedup}: {Latency-Aware}, Inline Data Deduplication for Primary Storage},
  booktitle = {10th {USENIX} Conference on File and Storage Technologies (FAST '12)},
  pages     = {1--14},
  year      = {2012}
}

@inproceedings{guerra2011cost,
  author    = {Jorge Guerra and Himabindu Pucha and Joseph Glider and Wendy Belluomini and Raju Rangaswami},
  title     = {Cost Effective Storage Using Extent Based Dynamic Tiering},
  booktitle = {9th {USENIX} Conference on File and Storage Technologies (FAST '11)},
  year      = {2011}
}

@inproceedings{zhang2010adaptive,
  author    = {Gong Zhang and Lawrence Chiu and Ling Liu},
  title     = {Adaptive Data Migration in {Multi-Tiered} Storage Based Cloud Environment},
  booktitle = {3rd {IEEE} International Conference on Cloud Computing (CLOUD '10)},
  pages     = {148--155},
  year      = {2010},
  publisher = {IEEE}
}

@article{ec-li2013erasure,
  author    = {Jun Li and Baochun Li},
  title     = {Erasure Coding for Cloud Storage Systems: A Survey},
  journal   = {Tsinghua Science and Technology},
  volume    = {18},
  number    = {3},
  pages     = {259--272},
  year      = {2013},
  publisher = {TUP}
}

@inproceedings{ec-plank2005t1,
  author    = {James S. Plank},
  title     = {{T1}: Erasure Codes for Storage Applications},
  booktitle = {4th {USENIX} Conference on File and Storage Technologies (FAST '05)},
  pages     = {1--74},
  year      = {2005}
}

@inproceedings{ec-li2019openec,
  author    = {Xiaolu Li and Runhui Li and Patrick P. C. Lee and Yuchong Hu},
  title     = {{OpenEC}: Toward Unified and Configurable Erasure Coding Management in Distributed Storage Systems},
  booktitle = {17th {USENIX} Conference on File and Storage Technologies (FAST '19)},
  pages     = {331--344},
  year      = {2019}
}

@inproceedings{ec-lee2016erasure,
  author    = {Ojus Thomas Lee and S. D. Madhu Kumar and Priya Chandran},
  title     = {Erasure Coded Storage Systems for Cloud Storage—Challenges and Opportunities},
  booktitle = {International Conference on Data Science and Engineering (ICDSE '16)},
  pages     = {1--7},
  year      = {2016},
  publisher = {IEEE}
}

@inproceedings{dict-kempa2018roots,
  author    = {Dominik Kempa and Nicola Prezza},
  title     = {At the Roots of Dictionary Compression: String Attractors},
  booktitle = {50th Annual {ACM} {SIGACT} Symposium on Theory of Computing (STOC '18)},
  pages     = {827--840},
  year      = {2018}
}

@article{dict-larsson2000off,
  author    = {N. Jesper Larsson and Alistair Moffat},
  title     = {{Off-line} {Dictionary-Based} Compression},
  journal   = {Proceedings of the IEEE},
  volume    = {88},
  number    = {11},
  pages     = {1722--1732},
  year      = {2000},
  publisher = {IEEE}
}

@software{fio,
  author       = {Jens Axboe},
  title        = {{Flexible I/O Tester}},
  year         = {2022},
  howpublished = {[Online]},
  url          = {https://github.com/axboe/fio}
}

@inproceedings{harnik2013zip,
  author    = {Danny Harnik and Ronen Kat and Dmitry Sotnikov and Avishay Traeger and Oded Margalit},
  title     = {To Zip or Not to Zip: Effective Resource Usage for {Real-Time} Compression},
  booktitle = {11th {USENIX} Conference on File and Storage Technologies (FAST '13)},
  pages     = {229--241},
  year      = {2013}
}

@inproceedings{harnik2016estimating,
  author    = {Danny Harnik and Ety Khaitzin and Dmitry Sotnikov},
  title     = {Estimating Unseen {Deduplication—from} Theory to Practice},
  booktitle = {14th {USENIX} Conference on File and Storage Technologies (FAST '16)},
  pages     = {277--290},
  year      = {2016}
}

@inproceedings{harnik2014fast,
  author    = {Danny Harnik and Ety Khaitzin and Dmitry Sotnikov and Shai Taharlev},
  title     = {A Fast Implementation of Deflate},
  booktitle = {Data Compression Conference (DCC '14)},
  pages     = {223--232},
  year      = {2014},
  publisher = {IEEE}
}

@inproceedings{abali2020data,
  author    = {Bulent Abali and Bart Blaner and John Reilly and Matthias Klein and Ashutosh Mishra and Craig B. Agricola and Bedri Sendir and Alper Buyuktosunoglu and Christian Jacobi and William J. Starke and others},
  title     = {Data Compression Accelerator on {IBM} {POWER9} and {z15} Processors: Industrial Product},
  booktitle = {47th Annual International Symposium on Computer Architecture (ISCA '20)},
  pages     = {1--14},
  year      = {2020},
  publisher = {IEEE}
}

@inproceedings{lu2013extending,
  author    = {Youyou Lu and Jiwu Shu and Weimin Zheng},
  title     = {Extending the Lifetime of {Flash-Based} Storage Through Reducing Write Amplification from File Systems},
  booktitle = {11th {USENIX} Conference on File and Storage Technologies (FAST '13)},
  pages     = {257--270},
  year      = {2013}
}

@inproceedings{lu2019ocstore,
  author    = {Youyou Lu and Jiacheng Zhang and Zhe Yang and Liyang Pan and Jiwu Shu},
  title     = {{OCStore}: Accelerating Distributed Object Storage with {Open-Channel} {SSDs}},
  booktitle = {39th {IEEE} International Conference on Distributed Computing Systems (ICDCS '19)},
  pages     = {271--281},
  year      = {2019},
  publisher = {IEEE}
}

@inproceedings{hotstorage,
author = {Huang, Jiawei and Li, Junru and Wang, Qing and Wen, Lijie and Lu, Youyou and Xu, Erci},
title = {CableCache: In-Network Request Deduplication for Key-Value Stores},
year = {2025},
isbn = {9798400719479},
publisher = {Association for Computing Machinery},
address = {New York, NY, USA},
url = {https://doi.org/10.1145/3736548.3737833},
doi = {10.1145/3736548.3737833},
booktitle = {Proceedings of the 17th ACM Workshop on Hot Topics in Storage and File Systems (HotStorage'25)},
pages = {86–92},
numpages = {7},
location = {Boston, MA, USA},
series = {HotStorage '25}
}

@inproceedings{cloudjump2,
author = {Chen, Zongzhi and Yang, Xinjun and Sha, Mo and Li, Feifei and Wang, Kang and Miao, Zheyu and Xu, Jie and Wang, Jianfeng and Wang, Sheng},
title = {CloudJump II: Optimizing Cloud Databases for Shared Storage},
year = {2025},
isbn = {9798400715648},
publisher = {Association for Computing Machinery},
address = {New York, NY, USA},
url = {https://doi.org/10.1145/3722212.3724431},
doi = {10.1145/3722212.3724431},
booktitle = {{ACM} {SIGMOD} International Conference on Management of Data (SIGMOD '25)},
pages = {336–349},
numpages = {14},
location = {Berlin, Germany},
series = {SIGMOD/PODS '25}
}

@article{zhang2017flashkv,
  title={FlashKV: Accelerating KV performance with open-channel SSDs},
  author={Zhang, Jiacheng and Lu, Youyou and Shu, Jiwu and Qin, Xiongjun},
  journal={ACM Transactions on Embedded Computing Systems (TECS)},
  volume={16},
  number={5s},
  pages={1--19},
  year={2017},
  publisher={ACM New York, NY, USA}
}

@inproceedings{oclu1,
  title={Ocvm: Optimizing the isolation of virtual machines with open-channel ssds},
  author={Liu, Zhe and Liao, Xiaojian and Li, Fei and Yang, Zhe and Lu, Youyou and Shu, Jiwu},
  booktitle={International Conference on Algorithms and Architectures for Parallel Processing},
  pages={416--432},
  year={2020},
  organization={Springer}
}

@article{oclu2,
  title={Mitigating synchronous I/O overhead in file systems on open-channel SSDs},
  author={Lu, Youyou and Shu, Jiwu and Zhang, Jiacheng},
  journal={ACM Transactions on Storage (TOS)},
  volume={15},
  number={3},
  pages={1--25},
  year={2019},
  publisher={ACM New York, NY, USA}
}



\end{document}